# Data-Driven Modeling of Amyloid-beta Targeted Antibodies for Alzheimer's Disease

Short Title: **Modeling Anti-Aβ Therapies in Alzheimer's Disease**

Kobra Rabiei[1]*, Jeffrey R. Petrella[2], Suzanne Lenhart[3], Chun Liu[4], P. Murali Doraiswamy[5–7], Wenrui Hao[1]

[1]Department of Mathematics, Penn State University, University Park, PA, USA.
[2]Department of Radiology, Duke University Health System, Durham, NC, USA.
[3]Department of Mathematics, University of Tennessee, Knoxville, TN, USA.
[4]Department of Applied Mathematics, Illinois Institute of Technology, Chicago, IL, USA.
[5]Duke Center for Applied Genomics and Precision Medicine, Duke University Health System, Durham, NC, USA.
[6]Department of Medicine, Duke University School of Medicine, Durham, NC, USA.
[7]Department of Psychiatry, Duke University School of Medicine, Durham, NC, USA.
*Corresponding author. Email: kxr5609@psu.edu.

**Abstract**

Alzheimer's disease (AD) is driven by the accumulation of amyloid-beta (Aβ) proteins in the brain, leading to memory loss and cognitive decline. While monoclonal antibodies targeting Aβ have been approved, optimizing their use to maximize benefits while minimizing side effects remains a challenge. This study develops a mathematical model to describe Aβ aggregation, capturing its progression from monomers to toxic oligomers, protofibrils, and fibrils using mass-action kinetics and coarse-grained modeling. The model is calibrated with experimental data, incorporating parameter estimation and sensitivity analysis to ensure accuracy. An optimal control framework is introduced to determine the best drug dosing strategy that reduces toxic Aβ aggregates while minimizing adverse effects, such as amyloid-related imaging abnormalities (ARIA). Results indicate that Donanemab achieves the greatest reduction in fibrils. This work provides a quantitative framework for optimizing AD treatment strategies, offering insights into balancing therapeutic efficacy and safety.

**Teaser**

A mathematical model optimizes Alzheimer's treatment by reducing toxic amyloid-beta while minimizing side effects.

**Introduction**

Alzheimer's disease (AD) is a devastating neurodegenerative disorder and the leading cause of dementia worldwide. Affecting over 47 million people globally, its prevalence is projected to rise significantly in the coming years, placing immense strain on healthcare systems (1). The disease is characterized by a progressive decline in cognitive function, including memory loss and diminished independence. These symptoms are closely linked to two hallmark pathological features: extracellular amyloid-beta (Aβ) plaques and intracellular neurofibrillary tangles composed of hyperphosphorylated tau protein. Together, these aggregates impair synaptic function, trigger chronic neuroinflammation, and lead to widespread neuronal loss (2–5).

Amyloid-beta (Aβ) aggregation plays a central role in the progression of Alzheimer's disease (AD). This process begins with the misprocessing of amyloid precursor protein (APP) by β- and γ-secretases, which results in the production of Aβ peptides. These peptides then aggregate into various forms, including toxic oligomers, fibrils, and insoluble plaques (6–8). Of these, toxic oligomers are considered the most neurotoxic, as they disrupt synaptic communication and are a major contributor to cognitive decline. In addition, secondary nucleation, a process in which fibrillar surfaces catalyze the formation of more toxic oligomers, has been identified as a key driver of amyloid propagation. This mechanism also represents a promising target for therapeutic intervention (9–11).

Given the complexity of Aβ aggregation, computational models have become essential tools for understanding the underlying mechanisms and exploring potential therapeutic interventions. Early models focused on the molecular dynamics of Aβ and its spread between neurons, using Smoluchowski equations and kinetic transport models. Later developments introduced spatial dynamics to account for the heterogeneous distribution of misfolded proteins throughout the brain. More advanced models, based on partial differential equations (PDEs), now capture the interactions among neurons, astrocytes, and microglia, providing valuable insights into possible therapeutic strategies. Recent innovations have further enhanced these models by incorporating clinical biomarkers and patient-specific parameters (12–15).

Experimental validation plays a crucial role in advancing these computational models. Structural studies offer detailed insights into molecular aggregation pathways, while kinetic experiments quantify the transition rates between oligomeric and fibrillar states under physiological conditions. These findings continually refine computational models, improving their predictive accuracy and ensuring their alignment with biological observations (2, 16, 17).

As computational models continue to advance, therapeutic strategies targeting Aβ are also evolving. Monoclonal antibodies, including Aducanumab, Donanemab, and Lecanemab, are designed to reduce plaque burden and slow cognitive decline. While some clinical trials have demonstrated promising results, challenges remain, such as inconsistent efficacy and adverse effects like amyloid-related imaging abnormalities (ARIA). In this context, computational models incorporating clinical and spatial data have become crucial tools for optimizing these therapies (18).

Building upon these advancements, optimal control techniques have increasingly been applied to design treatment strategies for AD, focusing on balancing therapeutic efficacy while minimizing side effects. Recent studies have explored the use of mathematical models incorporating optimal control to personalize anti-amyloid-beta therapy regimens (19, 20). These models leverage patient-specific data to optimize dosing strategies, providing deeper insights into potential therapeutic interventions and helping to tailor treatments for individual patients. However, most existing models focus on late-stage amyloid aggregation and plaque formation, while overlooking the detailed intermediate aggregation process. This gap limits their ability to evaluate early therapeutic interventions targeting toxic oligomers, which are now recognized as key drivers of neurotoxicity.

Additionally, current models often prioritize amyloid clearance while neglecting potential side effects, such as amyloid-related imaging abnormalities (ARIA), which remain a major challenge in clinical applications.

To address these gaps, this study develops a comprehensive mathematical framework that captures the full progression of Aβ aggregation, from monomers to higher-order aggregates, including toxic oligomers, protofibrils, and fibrils. By explicitly modeling intermediate states, our approach provides a detailed understanding of early aggregation dynamics and their role in AD pathology. Furthermore, we integrate an optimal control framework to optimize treatment regimens, not only focusing on reducing amyloid plaques but also minimizing toxic oligomers. Additionally, our model incorporates a clinically relevant side-effect function, ensuring that therapeutic strategies balance efficacy with safety.

In this paper, we develop a system of differential equations to model amyloid-beta (Aβ) aggregation, explicitly incorporating interactions across different molecular weight species. We apply mass-action kinetics for early-stage aggregation and a coarse-grained approach for larger aggregates, thereby balancing model detail and computational efficiency. Additionally, we integrate parameter estimation, identifiability analysis, and uncertainty quantification to ensure model robustness through validation with experimental data. Finally, we implement an optimal treatment framework aimed at minimizing both amyloid plaques and toxic oligomers while mitigating treatment side effects, using advanced optimization techniques.

## Results

We developed a mathematical model of amyloid-beta ($A\beta$) aggregation (Figure 1) using a system of ordinary differential equations to track its progression from monomers to fibrils. Sensitivity and identifiability analyses assess key parameters, while uncertainty quantification evaluates model reliability. Finally, we optimize treatment strategies to minimize fibril accumulation and toxic oligomers while reducing side effects, offering potential insights for Alzheimer's disease therapy.

### Parameter estimation

The values and ranges for 19 key parameters in the amyloid-beta (A$\beta$) aggregation model, including reaction rates, clearance rates, and saturation values, were estimated through literature review and steady-state analysis. Table 1 summarizes the results, distinguishing previously published values from those newly estimated for this model.

### Sensitivity Analysis

In this study, Sobol sensitivity analysis was used to assess the impact of model parameters on fibril dynamics $F(T)$, our model output at the final time $T = 21.147$ h, in line with experimental data (21). First-order, second order, and total sensitivity indices ($S_T$) were computed across different sample size $N$, using parameter ranges from Table 1. The initial conditions were set as $M(0) = 5\mu M$ and all other states at zero (21). The bar chart in fig. S1 highlights the most sensitive parameters for $N = 2^{15}$. To validate the sensitivity indices, Figure 2 displays the six most sensitive parameters across sample sizes ranging from $N = 2^4$ to $N = 2^{20}$. The x-axis represents $\log_2 N$, while the y-axis shows the Sobol total sensitivity indices of $F(T)$. The graph demonstrates convergence of these indices.

Additionally, fig. S2 presents the total sensitivity indices of $F(t)$ for these parameters at different time points, using a sample size of $N = 2^{15}$. Table S1 presents Sobol indices and confidence intervals for a sample size of $N = 2^{20}$. The confidence intervals are notably small, typically less than 10% of the corresponding Sobol indices, indicating high accuracy. Their narrow range further highlights the reliability of the estimates. Fig. S3 presents the first, second-order, and total sensitivity analyses, demonstrating consistency across different sensitivity measures. Node sizes represent the first-order Sobol indices ($S_i$), showing each parameter's direct impact on $F(T)$, while node borders indicate total sensitivity ($S_{T,i}$), capturing both direct and interaction effects. Edge thickness between nodes corresponds to second-order indices ($S_{i,j}$), highlighting parameter interactions.

We conducted a Sobol sensitivity analysis on our model under varying initial monomer concentrations, where $M(0)$ represents the initial concentration of the monomer, expressed in micromolar ($\mu M$) units
$$M(0) = 5, 4, 3.5, 3, 2.5, 2, 1.75, 1.5, 1.35, \text{ and } 1.1,$$
over their corresponding parameter ranges with $N = 2^{15}$. We obtained 10 total Sobol sensitivity indices for each parameter and plotted them in Figure 3. We set a sensitivity threshold at $10^{-5}$ to differentiate between sensitive and non-sensitive parameters. Specifically, parameters with average indices below this threshold are considered non-sensitive, while those with indices above it are deemed sensitive. More information and details about the total sensitivity of the model solution under different $M(0)$ values are provided in Supplementary Text B, as illustrated in fig S4.

**Data-driven modeling**

In this section, we calibrate our model using experimental data (21) and perform parameter identifiability and uncertainty quantification. First, using the parameter ranges specified in Table 1 and a detailed parameter estimation method, we identified a specific feasible region for certain parameters, dependent on the initial monomer concentration ($M(0) = 5, \ldots, 1.1 \mu M$). These results are summarized in Table 2.

Next, using normalized experimental data (fig. S5) and the Nelder-Mead optimization method, we determined an optimal parameter vector, $\theta^*$, for each of the 10 experimental observations. Practical identifiability classified parameters as identifiable or non-identifiable, with the latter requiring regularization for better model calibration. To handle non-identifiable parameters, we used a regularization strategy that prevented unstable values while keeping a good fit to the data. This approach gently guided poorly constrained parameters toward reasonable values by limiting large deviations from a reference set. As a result, the final optimized parameter vector, $\tilde{\theta}$, remains stable and reliable for each observation. After determining $\tilde{\theta}$ for 10 different initial concentrations of $M(0)$, we summarize the results in fig. S6. The detailed values of $\tilde{\theta}$ can be found in table S2 in Supplementary Text B. Finally, we assess model uncertainty arising from non-identifiable parameters and initial conditions by computing 95% confidence intervals for the model solutions. Fig. S7 displays the concentrations of all state variables for $M(0) = 2\mu M$, using the optimal parameter vector $\tilde{\theta}$. The shaded regions represent 95% confidence intervals, capturing variability from parameter and initial condition uncertainties. Figure 4 shows the 95% confidence intervals for $F(t)$ across nine additional initial monomer concentrations. The red curves represent model solutions with optimized parameters, while black dots correspond to experimental data. The green- shaded regions indicate the confidence intervals, capturing the uncertainty from non-identifiable parameters and initial conditions.

**Molecular Targets and Mechanisms of Anti-$A\beta$ Treatments**

Figure 1 schematically illustrates the molecular targets and mechanisms of Donanemab, Lecanemab, and Aducanumab in the amyloid-beta aggregation cascade. Table 3 summarizes the effectiveness of single and combined anti-amyloid-beta therapies in reducing fibril concentration ($F(t)$). Donanemab ($u_D$) was the most effective single-drug treatment, achieving over 85% fibril reduction across all initial monomer concentrations ($M(0)$). Lecanemab ($u_L$) and Aducanumab ($u_A$) produced moderate reductions of up to 25% and 27%, respectively. Combined therapies, particularly those including all three drugs ("All Drugs"), achieved the highest reductions, exceeding 87% in most cases.

Figures 5 and S8 show the optimal control function values for different initial conditions $M(0)$. The control functions for single-drug treatments (Donanemab: $u_D$; Aducanumab: $u_A$; Lecanemab: $u_L$) are compared with those from combined therapy scenarios. The solid and dashed lines corresponding to the single-drug and combined therapy are plotted in the same color for each initial condition. From the figure, we can see that when drugs are combined, treatments using Lecanemab ($u_L$) and Aducanemab ($u_A$) can be slightly delayed compared to when these drugs are used individually in the single treatment scenario. This suggests that combining the drugs allows for more flexibility in timing while maintaining effectiveness. The pattern for Donanemab ($u_D$) remains nearly unchanged between single and combined treatment scenarios. This is expected because Donanemab appears to be the dominant treatment, likely due to its higher maximum efficacy $U_{max}^D$, compared to the other drugs. Figure 6 illustrates the robustness of fibril reduction across different initial monomer concentrations. The 95% confidence intervals for fibril concentrations $F(t)$ following single-drug treatments highlight the model's reliability under varying initial conditions.

**Discussion**

This study develops a mathematical model of amyloid-beta ($A\beta$) aggregation, focusing on its transition from monomers to harmful fibrils. The model integrates differential equations, mass- action kinetics, and coarse-grained modeling to provide a detailed view of the aggregation process (Fig. 1). Sensitivity analysis, parameter estimation, and uncertainty quantification confirm the model's robustness and reliability, offering insights into Alzheimer's disease progression and potential treatment strategies. By identifying 19 key parameters that govern $A\beta$ aggregation, our parameter estimation and sensitivity analyses ensure that reaction rates, clearance mechanisms, and saturation effects align with experimental data. The consistency of Sobol sensitivity indices across different sample sizes further supports the model's stability. Additionally, sensitivity analysis across different initial monomer concentrations ($M(0)$) highlights the model's adaptability to biologically relevant conditions, making it useful for optimizing treatment strategies under various physiological settings. Calibrating th model with experimental data highlights the importance of data-driven methods in improving both predictive accuracy and interpretability. Using the parameter ranges from Table 1 and advanced estimation techniques, we identified feasible ranges for key parameters across different initial monomer concentrations ($M(0)$), as shown in Table 2. This calibration ensures that the model's predictions align well with experimental observations. Parameter identifiability analysis, using experimental data, showed that certain reaction rates and clearance parameters were identifiable. Uncertainty quantification confirmed the reliability of the parameter estimates, showing that aggregation pathway predictions remain consistent across experimental conditions. This stability is key to ensuring the model's reliable performance in therapeutic applications. By integrating experimental data with theoretical predictions, this data-driven approach enhances the precision and applicability of the $A\beta$ aggregation model, providing a strong framework for understanding Alzheimer's disease and optimizing treatments.

The treatment of AD has advanced with monoclonal antibodies like Donanemab, Lecanemab, and Aducanumab, each targeting different forms of amyloid-beta ($A\beta$) plaques. Donanemab specifically binds to $A\beta_{pE3}$, a modified form of fibrils found in plaques. While our model does not distinguish fibril subtypes like $A\beta_{pE3}$, it uses $F$, the final aggregated form, to reflect the aggregated species targeted by these therapies, consistent with their clinical function. Lecanemab targets protofibrils and plaque fibrils, while Aducanumab binds both soluble oligomers and fibrils. Al- though Aducanumab is no longer in clinical use due to an unfavorable risk-benefit profile, it is included in our model for comparative purposes. To evaluate treatment safety, we incorporate a side-effect function focused on ARIA-E, the most reported adverse effect, which shows a clear dose-dependent relationship with treatment. Although ARIA-H is not dose-dependent and often occurs in patients with existing ARIA-E, summing both ARIA-E and ARIA-H would over-estimate total ARIA incidence, as some patients experience both subtypes, and their overlap is not well defined. Therefore, we focus on ARIA-E to more accurately capture the dose-dependent nature of side effects. By simulating different treatment strategies, our model aims to optimize the balance between efficacy and safety. To study the effects of each drug, we incorporated them as control functions in our mathematical model. This helped us determine the best timing and dosage for treatment. The model also tracks all key stages of amyloid-beta aggregation, including protofibril, soluble oligomer, and fibril formation, giving a complete picture of how these therapies influence disease progression. Our results show that Donanemab is the most effective single treatment for reducing fibrils, achieving over 85% reduction across all initial conditions. When combined with Lecanemab and Aducanumab, the reduction improves slightly to over 87%, but this difference is not statistically significant. A key advantage of using Lecanemab and Aducanumab together is that their treatment timing can be more flexible, allowing for slower titration. Our study offers valuable insights into personalizing AD treatment. By simulating different starting conditions and treatment plans, our model shows how therapies can be customized for each patient. We also included a side-effect function based on clinical data, helping us balance the benefits of treatment with potential risks like amyloid-related imaging abnormalities (ARIA), ensuring safety alongside effectiveness. While our model provides a solid foundation for optimizing $A\beta$-targeting therapies, there's room for improvement. Incorporating patient-specific pharmacokinetics could make predictions more accurate and support better personalized strategies. Including additional side effects, such as immune responses or chronic inflammation, would give a more complete picture of treatment risks.

This approach could also be applied more broadly. Adding biomarkers for neuroinflammation or tau protein would help us better understand AD progression. Exploring combination therapies, like monoclonal antibodies paired with anti-tau drugs, could improve treatment outcomes. Plus, this framework could be adapted to other neurodegenerative diseases that involve protein aggregation.

**Materials and Methods**

The aggregation of amyloid-beta (A$\beta$) from monomers to fibrils is a complex and multi-step process involving several intermediate states. Initially, soluble A$\beta$ monomers, which are primarily in random coil or $\alpha$-helix conformations, associate to form dimers and trimers through non-covalent interactions (4). These small oligomers can further aggregate into larger fibrillar oligomers with $\beta$-sheet structures, leading to the formation of protofibrils, which are elongated precursors to mature fibrils (22). Some oligomers, however, remain as non-fibrillar and high molecular weight (HMW) oligomers, which are less structured but highly toxic (6). Eventually, the protofibrils elongate and stack to form insoluble, stable fibrils characterized by a cross-$\beta$ sheet structure. These fibrils accumulate as amyloid plaques in the brain, a hallmark of Alzheimer's disease (AD) (4). Understanding these intermediate stages is crucial for developing interventions to prevent or disrupt amyloid aggregation (22). In Figure 1, we summarize the $A\beta$ aggregation process (23) and define the following variables to categorize the various species of A$\beta$ aggregates:

- $M$: Monomers
- $D$: Dimers
- $T$: Trimers
- $P$: Fibrillar oligomers and Protofibrils
- $N$: Non-fibrillar oligomers and High molecular weight (HMW) oligomers
- $F$: Fibrils

We propose a system of differential equations to develop a mathematical model that describes the dynamics of A$\beta$ aggregation, as illustrated in Figure 1. Then we model this process by following two principal parts below:

1- **Mass Action Kinetics**: This traditional approach will be applied to the formation and inter- action dynamics of the lower molecular weight species $M$, $D$, and $T$. The mass action law provides a framework to quantify how these species combine or convert from one form to another over time.
2- **Coarse-Grained Modeling:** For the higher molecular weight complexes $P$, $N$, and $F$, a coarse-grained approach is adopted. This method allows us to represent the aggregation behavior of these larger complexes more abstractly, which is useful in capturing the essential features of their dynamics without detailing every underlying interaction.

This newly developed model in this paper aims to comprehensively describe the aggregation pathways and kinetics of A$\beta$ species, facilitating deeper insights into the mechanisms driving the progression of amyloid-related diseases and exploring the efficacy of possible treatments.

**The mass action model**

To accurately model the kinetics of Aβ aggregation, we consider the fundamental chemical reactions among monomers ($M$), dimers ($D$), and trimers ($T$). These reactions are pivotal in understanding the aggregation pathway that leads to amyloid fibril formation. Below, we present the primary reactions and their corresponding rate constants:

$$2M \underset{k_{21}}{\overset{k_{12}}{\rightleftharpoons}} D, \quad 3M \underset{\epsilon_1}{\overset{k_{13}}{\rightleftharpoons}} T, \quad M + D \underset{\epsilon_2}{\overset{k_{23}}{\rightleftharpoons}} T. \tag{1}$$

Using the general formula of mass action provided in Supplementary Text A, we can derive the following equations for monomers, dimers, and trimers that describe these chemical reactions.

**The equation of Monomers**

The dynamics of monomer concentration are influenced by various processes, presented in the following differential equation

$$\frac{d\,[M]}{dt} = \widehat{\mu_1} - k_{12}[M]^2 + k_{21}[D] - k_{13}[M]^3 + \epsilon_1[T] - k_{23}[M][D] + \epsilon_2[T] - \mu_1[M]. \tag{2}$$

There are several key terms in this equation including: $\widehat{\mu_1}$ represents the production rate of monomers; $k_{12}[M]^2$ describes the aggregation of monomers to form dimers; $k_{21}[D]$ models the dissociation of dimers into monomers; $k_{13}[M]^3$ represents the aggregation of monomers to form trimers; $\epsilon_1[T]$ means the conversion of trimers back to monomers; $k_{23}[M][D]$ describes the interaction between monomers and dimers to form trimers; $\epsilon_2[T]$ is the conversion of trimers back to monomers and dimers; and $\mu_1[M]$ is the clearance or degradation of monomers.

**The equation of Dimers**

The rate of change in dimer concentration is given by:
$$\frac{d\,[D]}{dt} = k_{12}[M]^2 - k_{21}[D] - k_{23}[M][D] + \epsilon_2[T] - \mu_2[D], \tag{3}$$

Where $\mu_2[D]$ represents the degradation rate of dimers.

**The equation of Trimers**

Finally, the rate of change in trimer concentration is described by:
$$\frac{d\,[T]}{dt} = k_{13}[M]^3 + k_{23}[M][D] - \epsilon_1[T] - \epsilon_2[T] - \mu_3[T], \tag{4}$$

Where $\mu_3[T]$ is the clearance of trimer.

**The Coarse-Grained Model**

To model the dynamics of larger molecular weight species in Aβ aggregation, we employ a coarse-grained approach as outlined in the following reaction scheme:

$$\alpha D \to P, \qquad \beta T \to N, \; P + N \to F \qquad (5)$$

This model simplifies the complex interactions into direct pathways leading to the formation of larger complexes, thereby enabling a focus on the essential aspects of their behavior.

We introduce the coefficients $\alpha$ and $\beta$, which represent the numbers of dimers (D) and trimers (T) to form fibrillar oligomers and protofibrils (P), as well as non-fibrillar oligomers and high molecular weight oligomers (N), respectively.

In this case, we introduce $\tilde{D} = \alpha D$ and $\tilde{T} = \beta T$, the formations of (P) and (N) are directly tied to the concentrations of $\tilde{D}$ and $\tilde{T}$, respectively.

**Equation for Fibrillar Oligomers and Protofibrils**

The concentration of $P$ is described by the following logistic growth model:

$$\frac{d[P]}{dt} = \widetilde{\lambda_1}[\tilde{D}](\widetilde{K_1} - \tilde{D}) - \mu_4[P] = \widetilde{\lambda_1}[\alpha D](\alpha K_1 - \alpha D) - \mu_4[P],$$

which implies that
$$\frac{d[P]}{dt} = \lambda_1[D](K_1 - [D]) - \mu_4[P] \qquad (6)$$
where $\lambda_1 = \alpha^2 \widetilde{\lambda_1}$ and $K_1$ represents the carrying capacity of $D$.

**Equation for Non-fibrillar Oligomers and High Molecular Weight Oligomers**

Similarly, the dynamics of $N$ are modeled by:
$$\frac{d[N]}{dt} = \lambda_2[T](K_2 - [T]) - \mu_5[N] \qquad (7)$$
where $K_2$ is the saturation limit of $T$, reflecting the maximum feasible concentration.

**Equation for Fibrils**

The formation of fibrils $F$ is influenced by both $P$ and $N$, with the following Michaelis-Menten Kinetics:

$$\frac{d[F]}{dt} = \lambda_3 \frac{[P]}{1+[N]/K_3} - \mu_6[F]. \qquad (8)$$

**Parameter Estimation**

Estimating parameter values is essential for ensuring that mathematical models accurately reflect observed data. Whenever possible, we use literature-based values, as they are well-documented and reliable. When unavailable, a range of values is considered. For reactions covered in previous studies, we adopt existing parameter ranges, while for new reactions unique to our model, we estimate parameters using steady-state concentrations. Throughout this paper, aggregate concentrations are reported in molar ($M$) or micromolar ($\mu M$) units.

Parameters for the mass action model

In this subsection, we consider the monomer-dimer-trimer process described by equations (2), (3), and (4). The law of Mass Action is utilized and then taking into account existing

literature and the steady-state concentrations, we determine the parameter ranges as follows:

**Parameters in Eq. (2):**

From literature: We take the range of $k_{12}$ as $800 - 1180$ and $k_{23}$ as $33 - 43$ from (24) since our model closely resembles the monomer-dimer-trimer process outlined by (24). We take the range of the death rate, $\mu_1$, as $10^{-5}$ to $10^{-3}$ from (25) (consistent with $\mu_1 = 10^{-5}$ in (26)), and the birth rate, $\widehat{\mu_1}$, as $(10^{-14}, 10^{-12})$ from (27) (consistent with $\widehat{\mu_1} = 10^{-12}$ in (26)). Since the models in literature lack considering the reverse of certain reactions, we take $\epsilon_1$ within the range $(10^{-13}, 10^{-12})$.

From steady states: Reactions among monomer, dimer, and trimer adhere to the mass action rule and constitute a closed system, the concentrations of dimer and trimer at steady state can reach $10^{-6}$. Consequently, the concentration of monomers should decrease; therefore, we assume that the concentrations of $M$ and $D$ at the steady-states situation are $[M^\infty] \approx 10^{-9}$ and $[D^\infty] \approx 10^{-6}$. By considering on the steady-state of the reaction between $M$ and $D$ only, we have

$$k_{21} = \frac{k_{12}[M^\infty]^2}{[D^\infty]} = (8 \times 10^{-10}, 1.18 \times 10^{-9}).$$

**Parameters in Eq. (3)**

In this equation, the value of $k_{23}$ remains the same as previously defined. We have already set $k_{12}$ and $k_{21}$ in the previous equations to maintain the steady state. By using the steady-state condition, we have $\epsilon_2[T] \approx k_{12}[M]^2$, which implies $\epsilon_2 \approx 10^{-9}$. Since the conversion of the trimer back to monomer and dimer is negligible (8), we choose a range of $(10^{-11}, 10^{-9})$ for $\epsilon_2$. We choose the same range for $\mu_2$ as for $\mu_3$ since the clearance rates of the dimer and trimer are similar.

**Parameters in Eq. (4)**

By considering the steady state of Eq. (4) and neglecting $\epsilon_2[T^\infty]$ which is nearly zero, we have:

$$k_{13} = \frac{\epsilon_1[T^\infty]}{[M^\infty]^3} \qquad \text{and} \qquad \mu_3 = \frac{k_{23}[M^\infty][D^\infty]}{[T^\infty]}.$$

By choosing $\epsilon_1 \approx 10^{-12}$, we have $k_{13} \approx 10^9$. Similarly, by setting $k_{23} \approx 40$, we determine $\mu_3 \approx 10^{-8}$.

**The Coarse-Grained model**

In this subsection, we attempted to establish a range for the parameters appearing in Equations (6), (7), and (8). Since these parameters are not directly reported or utilized in other literature, we initially identified an acceptable range for them.

From Literature: The degradation rates $\mu_4$, $\mu_5$, and $\mu_6$ are reported in and range from $10^{-7}$ to $10^{-3}$.

**Parameters in Eq. (7)**

The concentration of dimers can reach around $10^{-6}$, we take the carrying capacity of $D$, $K_1$, to be in a range of $(10^{-7}, 4 \times 10^{-6})$. Since $P$ is formed by the accumulation of many dimers, $[P^\infty]$ is much greater than $[D^\infty]$. On the other hand, $P$ combines with $N$ to form fibrils, the concentration of which at a steady state is around $10^{-2}$ according to experimental data. Therefore, we assume $[P^\infty] = (10^{-4}, 10^{-3})$. Additionally, we assume $\mu_4 = (10^{-4}, 10^{-3})$ and have $[D^\infty](K_1 - [D^\infty])$ vary between $10^{-14}$ and $10^{-12}$, we determine $\lambda_1$ based on the steady state,

$$\lambda_1 = \frac{\mu_4 [P^\infty]}{[D^\infty](K_1 - [D^\infty])} = (10^5, 10^7).$$

**Parameters in Eq.(7)**

Trimers are formed from monomers and dimers, with the concentration around $10^{-6}$. Since $T$ arises from the combination of these species, its steady-state concentration, $[T^\infty]$, is greater than that of dimers, $[D^\infty]$. To reflect this, we set the carrying capacity $K_2$ within the range $(10^{-6}, 4 \times 10^{-6})$, ensuring $K_2$ has a larger lower bound compared to $K_1$. This setting maintains the inequality $10^{-12} \leq [T^\infty](K_2 - [T^\infty]) \leq 10^{-11}$. Since $[D^\infty](K_1 - [D^\infty])$ is approximated up to $10^{-12}$, we use $[T^\infty](K_2 - [T^\infty]) \approx 10^{-11}$ to estimate a range for $\lambda_2$. Similar to $P$, we set $[N^\infty]$ between $10^{-4}$ and $10^{-3}$ and by selecting $\mu_5 \approx 10^{-4}$ at steady state, $\lambda_2$ is estimated as follows:

$$\lambda_2 = \frac{\mu_5 [N^\infty]}{[T^\infty](K_2 - [T^\infty])} = (10^3, 10^4).$$

**Parameters in Eq. (8)**

In this equation, $K_3$ represents the carrying capacity for $N$, and thus we take $K_3 = [N^\infty] \approx 10^{-3}$. In the steady-state equation, we have

$$\frac{1}{2}\lambda_3 [P^\infty] - \mu_6 [F^\infty] = 0.$$

By setting $[P^\infty] = 10^{-4}$, and assuming $\mu_6 = (5 \times 10^{-5}, 10^{-4})$, we have $\lambda_3$ be within the range $[0,1]$. It is important to note that in this equation, we consider a normalized $F$ with $[F^\infty] = 1$ which has been widely considered in the literature (21). In Table 1, we summarize the values and ranges for the parameters of equations (2)–(8).

**Sensitivity Analysis**

In this section, we employ Sobol sensitivity analysis, a variance-based methodology derived from the analysis of variance (29). The Saltelli sampling approach, a derivative of Quasi-Monte Carlo methods, was employed for effective Sobol sensitivity analysis (30). For simplicity, we denote our model as

$$\frac{d\mathbf{x}}{dt} = \mathbf{f}(\mathbf{x}, \boldsymbol{\theta}),$$

where $\mathbf{x} = [M, D, T, P, N, F]$ represents the state variables, $\boldsymbol{\theta}$ denotes all the parameters, and $\mathbf{f}$ is the right-hand side of the system of differential equations. Initially, we generated $N$ samples for each parameter, forming the basis of our parameter space exploration. For each parameter, $\theta_i$, we created two matrices: one varying $\theta_i$ while keeping other parameters fixed, and another varying all parameters except $\theta_i$. This approach helps assess the individual and combined effects of parameters, essential for first and second-order Sobol sensitivity analyses. Using $N(2n_p + 2)$ sample points, where $N$ is the base sample size and $n_p$ is the number of parameters, the Saltelli method improves the convergence rate, leading to more precise results with fewer samples compared to traditional sampling.

Data-driven modeling

In our study, we use experimental data collected following a well-established protocol for the aggregation of beta-amyloid ($\beta$-amyloid) peptides. These measurements were performed using Thioflavin T, a dye that lights up more as aggregate levels increase. The data collection and subsequent analyses adhere strictly to the methods outlined in (9). Quality control measures are applied to ensure the reliability and reproducibility of the data. As highlighted in (21), this includes controls over experimental conditions such as protein purity and the validation of assumptions like the proportionality of dye fluorescence to aggregate concentrations. Following these rigorous standards is crucial not only for data accuracy but also for the validity of any conclusions drawn from the kinetic analysis. This meticulous approach allows us to extract meaningful data of fibrils into the mechanisms of peptide aggregation, underpinning the scientific value of our modeling efforts. This data includes the longitudinal concentration of fibrils, denoted as $F$ in our model, for 10 different initial monomer concentrations. According to literature (9-11, 21), it is a common approach to use normalized data for calibration. Therefore, we first normalize the available data set and map it between 0 and 1 such that the process starts from 0 as the initial value of $F$ and reaches a steady state at $F(T) = 1$. Fig. S5 shows both the raw and normalized concentrations of $F(t)$ for different initial conditions of $M(0)$.

We outlined our data-driven approach in Figure 7. The feasible parameter range is summarized in Table 1, and the sensitivity analysis identifies the non-sensitive parameters based on Figure 3. Thus, we fix these non-sensitive parameters and choose the lower bound of these parameters as their fixed values, as given in Table S3. Following these preprocessing procedures, in this section, we will calibrate our model using experimental data and perform parameter identifiability and uncertainty quantification.

Feasible parameter ranges for different initial monomer concentrations

Since some parameters, such as the carrying capacities for dimers and trimers, $K_1$ and $K_2$, highly depend on the initial monomer values, $M(0)$, we divide the initial values into two groups: $M(0) > 2\mu M$ and $M(0) \leq 2\mu M$. The parameter ranges listed in Table 1 are suitable for initial values in the first group. However, for $M(0) \leq 2\mu M$, it is necessary to reduce the upper bound of the $K_1$ and $K_2$ ranges from $4 \times 10^{-6}$ to $2 \times 10^{-6}$. As a result, the range of $\lambda_2$ should also be updated.

We recall that in Section **Parameters in Eq.(7)**, we used the inequality $10^{-12} \leq [T^\infty](K_2 - [T^\infty]) \leq 10^{-11}$ and approximated this term with $10^{-11}$ to find a range for $\lambda_2$ as $(10^3, 10^4)$. Changing the upper bound of $K_2$ from $4 \times 10^{-6}$ to $2 \times 10^{-6}$ requires using $10^{-12}$ as the closer approximation value for $[T^\infty](K_2 - [T^\infty])$, which makes $\lambda_2$ fall within the range $(10^4, 10^5)$. Thus, we summarize the three parameter values in Table 2.

Model calibration based on experimental data

The ranges specified in Tables 1 and 2 define a feasible region within the parameter space for $M(0) = 5,\ldots,1.1,$micromolar$(\mu M)$. To find an optimal parameter vector, $\theta^*$, By using the normalized experimental data and Nelder-Mead optimization method (31), we solve the following optimization problem to obtain the optimal values for sensitive parameters:

$$\theta^* = \operatorname*{argmin}_{\theta} \sum_{j=1}^{N} (F(t_j; \boldsymbol{\theta}) - F_j)^2, \tag{9}$$

where $F(t_j; \boldsymbol{\theta})$ represents the solution of Eq. (8) at time $t_j$ and $F_j$ is the normalized experimental data at $t_j$. This approach enables us to determine the parameter values, achieving an optimal match between the observed data and the simulated solution to $F$.

Parameter practical identifiability analysis

Next, we perform the practical identifiability analysis for all the parameters. As discussed in **Supplementary Text C**, the parameter identifiability is related to the sensitivity matrix defined as

$$S = \begin{bmatrix} s_1(t_1) & s_2(t_1) & \ldots & s_q(t_1) \\ s_1(t_2) & s_2(t_2) & \ldots & s_q(t_2) \\ \vdots & \vdots & \ddots & \vdots \\ s_1(t_N) & s_2(t_N) & \ldots & s_q(t_N) \end{bmatrix} \tag{10}$$

where $s_k(t_j) = \frac{\partial y(t_j, \theta^*)}{\partial \theta_k}$ and $t_j$ is corresponding to the given data $y_j$. According to **Supplementary Text C**, we decompose the $S^T S$ matrix as

$$S^T S = U \Lambda U^T, \tag{11}$$

where $U$ is an $n_p \times n_p$ matrix of eigenvectors, and $\Lambda$ is a diagonal matrix corresponding to the eigenvalues, e.g. $\lambda_1 > \lambda_2 > \cdots > \lambda_r > 0$, where $r$ is the rank. Thus, we can split the non-zero and zero eigenvalues of $S^T S$ as

$$S^T S = [U_1 \quad U_2] \begin{bmatrix} \Lambda_1 & 0 \\ 0 & 0 \end{bmatrix} \begin{bmatrix} U_1^T \\ U_2^T \end{bmatrix}, \tag{12}$$

where $\Lambda_1$ contains non-zero eigenvalues $\lambda_1, \cdots, \lambda_r$. In this case, we can decompose the parameters into two groups:

- **Identifiable Parameters**: These are defined as $U_1^T \theta$, corresponding to the non-zero eigenvalues. Specifically, as shown in **Supplementary Text C**, we can express

$$\nabla L_\epsilon(\theta) = \Lambda_1 U_1^T(\theta - \theta^*) + O(\epsilon) + O(\|\theta - \theta^*\|^2) = 0,$$

which means that we can solve $U_1^T \theta$ based on the optimization problem, making them identifiable.

- **Non-identifiable Parameters**: These are defined as $U_2^T \theta$, corresponding to the kernel of $S^T S$. These parameters cannot be identified by solving the optimization problem directly. To address this, we regularize the optimization problem by introducing a regularization term.

In practice, we determine the non-zero eigenvalues by setting a tolerance, e.g., $tr = 10^{-5}$, to set the identifiable and non-identifiable parameters. Eigenvectors associated with large eigenvalues $\lambda_i > tr$ indicate directions in the parameter space with significant variance, meaning these directions are well-determined by the data and are considered *identifiable*. Conversely, eigenvectors corresponding to small or near-zero eigenvalues, e.g., $\lambda_i < tr$, suggest that changes in these directions hardly affect the output. This implies that the data does not provide sufficient information to uniquely determine parameters in these directions. Therefore, any linear combination of parameters corresponding to these eigenvectors remains uncertain or *non-identifiable*.

Therefore, we attempt to enhance our calibration by considering the following minimization problem with an added regularization term:

$$\tilde{\theta} = \operatorname{argmin}\left(\sum_{j=1}^N (y(t_j, \theta) - y_j)^2 + \lambda \|U_2^T \theta - U_2^T \theta^*\|^2\right), \quad (13)$$

Here, the cost function consists of two parts. The first term measures the sum of squared differences between model predictions $y(t_j, \theta)$ at time $t_j$ and observed data $y_j$. This term ensures that the model fits the data well and the regularization term is specifically designed to handle the issue of non-identifiability. The regularization term penalizes the difference between the projected values of the new parameter vector $\theta$ and the reference vector $\theta^*$ onto the subspace defined by $U_2$. The regularization parameter $\lambda$ controls the balance between fitting the model to the data and penalizing large deviations from $\theta^*$. After optimizing the parameter values for 10 different initial concentrations of $M(0)$, we summarized the statistics of these values in fig. S6. The detailed values of $\tilde{\theta}$ are summarized in Table S2 of **Supplementary Text D**.

**Unceratiany quantification**

In this part, we assess the model uncertainty introduced by both the non-identifiable parameters and the initial condition. Specifically, we first perturb the initial condition of $F$ by $F(0) = \epsilon$, where $\epsilon$ follows a normal distribution, $\mathcal{N}(0, 0.05)$. This perturbation quantifies the model uncertainty introduced by the initial value of $F(0)$. Secondly, to address uncertainties introduced by the non-identifiable parameters, we define a perturbation vector $\varXi$ whose elements follow a normal distribution. The model parameters are then perturbed by:

$$\bar{\theta} = \tilde{\theta} \times (1 + (U_2^T)^{-1} \varXi) \quad (14)$$

where $(U_2^T)^{-1}$ is the pseudoinverse of $U_2^T$, projecting the perturbations onto the subspace defined by the non-identifiable directions.

By randomly generating 1000 perturbation vectors, we solved the model using the perturbed parameters and initial conditions and computed the 95% confidence interval for these solutions. In fig. S7, the concentrations of all state variables for the initial condition $M(0) = 2\mu M$ are plotted using the optimal parameter vector $\tilde{\theta}$, along with their respective confidence intervals. Additionally, Figure 4 presents the confidence intervals, model solutions $F$, and observations for other initial conditions $M(0)$. This comprehensive analysis allows us to assess the impact of parameter uncertainties on the simulated solution's reliability under varied conditions influenced by the non-identifiable parameters.

**Optimal treatment studies of Anti-Amyloid-beta Monoclonal Antibodies**

In recent years, significant advancements have been made in the therapeutic landscape of AD, particularly with the development of monoclonal antibody drugs such as donanemab, lecanemab, and aducanumab. These drugs aim to address the underlying amyloid pathology associated with AD by targeting amyloid-beta (A$\beta$) aggregates in the brain, as schematically illustrated in Figure 1.

- **Donanemab**: This drug specifically targets fibrillar amyloid-beta plaques and fibrils. Donanemab binds to particular epitopes on the fibrils, facilitating their clearance through immune-mediated mechanisms, promoting the removal of these aggregates (32).

- **Lecanemab**: Lecanemab primarily targets A$\beta$ protofibrils, which are intermediate aggregates in the pathway to form insoluble fibrils. This action helps prevent the progression to larger, more stable amyloid structures (32).

- **Aducanumab**: Aducanumab has a broader target range, including soluble oligomers and insoluble fibrils. By binding to these aggregates, aducanumab activates microglial cells, enhancing the phagocytosis and clearance of amyloid plaques (33).

These therapeutic strategies focus on targeting various stages of the amyloid-beta aggregation cascade to mitigate AD progression and improve patient outcomes. In our model, these anti-amyloid-beta therapies are introduced as control functions $u_D(t)$ (Donanemab), $u_L(t)$ (Lecanemab), and $u_A(t)$ (Aducanumab), each representing the respective drug's effect on reducing amyloid burden. By incorporating these control functions into the state equations, we derive the following equations for $P$, $N$, and $F$, while the equations for monomers, dimers, and trimers Eqs. (2-4) remain unchanged. Together, these equations form the state system for the optimal control problem.

$$\begin{aligned}
\frac{d[P]}{dt} &= \lambda_1[D](K_1 - [D]) - \mu_4[P] - u_L(t)[P] - u_A(t)[P] \\
\frac{d[N]}{dt} &= \lambda_2[T](K_2 - [T]) - \mu_5[N] - u_A(t)[N] \\
\frac{d[F]}{dt} &= \lambda_3\left(\frac{[P]}{1+[N]/K_3}\right) - \mu_6[F] - u_D(t)[F] - u_L(t)[F] - u_A(t)[F]
\end{aligned} \quad (15)$$

The optimal anti-amyloid-beta treatment seeks to reduce fibrils $F(t)$ and toxic aggregated species $N(t)$ while minimizing side effects. The objective function, which accounts for

the effects of three treatments—Donanemab, Lecanemab, and Aducanumab—is defined as follows:

$$J(u_D, u_L, u_A) = \gamma F(T) + \int_0^T (j_1 N(t) + j_2 F(t) \\ + \frac{1}{2} g(t) \left[ \left(\frac{u_D(t)}{U_{\max}^D}\right)^2 + \left(\frac{u_L(t)}{U_{\max}^L}\right)^2 + \left(\frac{u_A(t)}{U_{\max}^A}\right)^2 \right] \\ + c_D S_D(u_D) F(t) + c_L S_L(u_L) F(t) + c_A S_A(u_A) F(t)) \, dt. \tag{16}$$

We seek to find an optimal control vector $(u_D^*, u_L^*, u_A^*)$ to minimize our objective function,

$$J(u_D^*, u_L^*, u_A^*) = \min_{(u_D, u_L, u_A)} J(u_D, u_L, u_A).$$

The cost function addresses several objectives related to effective and safe treatment strategies. The first term, $\gamma F(T)$, minimizes the concentration of fibrils $F(T)$ at the end of the treatment period $T$, targeting the reduction of amyloid plaques. The integral terms $j_1 N(t)$ and $j_2 F(t)$ measure the total accumulation of toxic oligomers $N(t)$ and fibrils $F(t)$ over time. The quadratic terms $\left(\frac{u_D(t)}{U_{\max}^D}\right)^2$, $\left(\frac{u_L(t)}{U_{\max}^L}\right)^2$, and $\left(\frac{u_A(t)}{U_{\max}^A}\right)^2$, scaled by the time-dependent function $g(t)$, act as regularization terms that limit fluctuations and large magnitudes in the control variables $u_D(t)$, $u_L(t)$, and $u_A(t)$. The decay function $g(t)$ modifies the contribution of these terms over time, allowing the model to incorporate changes in treatment intensity or dosage.

The final terms $c_D S_D(u_D) F(t)$, $c_L S_L(u_L) F(t)$, and $c_A S_A(u_A) F(t)$ capture the potential side effects of each drug. Side effects are influenced by two factors: high drug dosage and the presence of larger amyloid plaques. This interaction is represented by the product of $S(u)$, the side effect function of the drug, and $F(t)$, the fibril concentration.

Our model focuses on ARIA-E, a subtype of amyloid-related imaging abnormalities (ARIA), as the key measure of drug side effects because it is both common and closely linked to dosage. ARIA-E, which causes brain swelling or fluid buildup, tends to increase with higher drug doses, making it a useful marker for evaluating treatment risks. The controls are $L^\infty(0, T)$ functions with constraints:

$$0 \leq u_D(t) \leq U_{\max}^D, \quad 0 \leq u_L(t) \leq U_{\max}^L, \quad 0 \leq u_A(t) \leq U_{\max}^A, \quad \forall t \in [0, T].$$

To solve this optimal control problem, we will analyze the effects of each drug individually. By isolating the impact of each treatment, Donanemab, Lecanemab, and Aducanumab, we aim to understand their specific contributions to therapeutic results.

### Analysis of Donanemab Treatment

Now, we consider $u_D$ first and set $u_L = u_A = 0$ in Eq. (15). In addition, we suppose that $g(t) = \alpha e^{-\beta t}$ and the cost function to be minimized becomes,

$$J(u_D) = \gamma F(T) + \int_0^T \left( j_1 N(t) + j_2 F(t) + \frac{1}{2} \alpha e^{-\beta t} \left(\frac{u_D(t)}{U_{\max}^D}\right)^2 + c_D S_D(u_D) F(t) \right) dt.$$

To derive the adjoint equations for this optimal control problem, we first introduce the Hamiltonian $H$ (34):

$$H\big(x(t), u(t), \Lambda(t)\big) = L\big(x(t), u(t)\big) + \Lambda(t)^\top f\big(x(t), u(t)\big),$$

where $L\big(x(t), u(t)\big)$ represents the Lagrangian of the system, the cost function integrand, and $\Lambda(t)$ is the vector of adjoint variables. The term $f\big(x(t), u(t)\big)$ represents the right-hand side of the system of ODEs, which includes the state variables $M, D, T, P, N, F$, along with the control functions. It is important to clarify that, although the ODE system was initially expressed as

$$\frac{dx}{dt} = f(x(t), \theta),$$

where $\theta$ represents the parameters, once these parameters are calculated and fixed, we include the control $u(t)$ in the optimal control problem. Therefore, the right-hand side is simplified to $f\big(x(t), u(t)\big)$. To focus specifically on the effects of Donanemab, we set $u_L = 0$ and $u_A = 0$.

Using Pontryagin's Maximum Principle, the adjoint equations for this scenario are derived and detailed in Eq. (S13) in **Supplementary Text E**. These equations follow the general form:

$$\frac{d\Lambda(t)}{dt} = -\frac{\partial H}{\partial \mathbf{x}},$$

where $\Lambda(t) = [\lambda_M, \lambda_D, \lambda_T, \lambda_P, \lambda_N, \lambda_F]^\top$ is the vector of adjoint variables. The boundary conditions for the adjoint variables are specified as:

$$\Lambda(T) = [0,0,0,0,0,\gamma]^\top,$$

where $\gamma$ is a constant, and the adjoint variables corresponding to the state variables $M$, $D$, $T$, $P$, and $N$ are zero at the final time $T$, with only $\lambda_F = \gamma$, being nonzero. To estimate the maximum dosage of Donanemab ($U_{\max}^D$) and its associated side effects ($S_D$), we use clinical data and define them as follows:

**Maximum Dosage** ($U_{\max}^D$): Represents the maximum clearance rate of amyloid-beta plaques induced by Donanemab. $U_{\max}^D = \frac{43}{7}\mu_6$ is calculated based on the reduction of amyloid plaque levels observed in clinical studies (35).

**Side Effects** ($S_D$): Captures the probability of Amyloid-Related Imaging Abnormalities (ARIA-E) associated with Donanemab. $S_D$ is modeled as proportional to the administered dosage $u_D(t)$ relative to $U_{\max}^D$:

$$S_D(u_D) = 0.221 \frac{u_D(t)}{U_{\max}^D}.$$

Further details regarding these estimations are provided in **Supplementary Text F**.

## Optimal Control Problem with Estimated Parameters

In this section, we set $j_1 = j_2 = c_D = 1$, $\alpha = 2$, and $\beta = 2 \times 10^{-4}$. The maximum dosage $U_{\max}^D$ and the side effect function $S_D(u_D)$ are used as previously estimated. The optimal control problem is then solved using the forward-backward sweep algorithm with boundary conditions. The algorithm iteratively adjusts the control $u_D(t)$ to reduce the cost function and achieve convergence to an optimal solution.

To solve the optimal control problem, we begin by initializing the control vector $u_D(t)$ to zero. The system of state differential equations is then solved forward in time using the initial conditions and the Runge-Kutta method to compute the state variables. These state variables are subsequently used to solve the adjoint equations backward in time, starting from the boundary conditions, to calculate the adjoint variables (**Algorithm 1**). The optimal control $u_D(t)^*$ is derived from the Hamiltonian condition on the interior of the control set:

$$\frac{\partial H}{\partial u_D} = 0 \Rightarrow \frac{0.221 c_D F}{U_{\max}^D} + \alpha e^{-\beta t} \frac{u_D(t)}{(U_{\max}^D)^2} - \lambda_F F = 0.$$

Taking the bounds into account, $0 \leq u_D(t) \leq U_{\max}^D$, leads to the following updated control:

$$u_{\text{new}}^D(t) = \min\left( U_{\max}^D, \max\left( 0, (U_{\max}^D)^2 \, \frac{\lambda_F F - \frac{0.221 c_D F}{U_{\max}^D}}{\alpha e^{-\beta t}} \right) \right).$$

The control is iteratively updated for stability and convergence using the following formula and this iterative process continues until $u_D(t)$ converges to a stable solution.

$$u_D(t) = 0.5\, u_D(t) + 0.5\, u_{\text{new}}^D(t).$$

We solve the optimal control system for each scenario by using 10 different initial conditions for the monomer and their corresponding parameters, as determined in the previous sections. In Table 3, the row corresponding to treatment $D$ presents the percentage reduction in fibril concentration, calculated as

$$\frac{F(T) - F_u(T)}{F(T)} \times 100.$$

Here, $F(T)$ represents the fibril concentration at the final time without any treatment ($u_D = 0$), while $F_u(T)$ denotes the fibril concentration at the same time in the presence of treatment ($u_D \neq 0$).

## Analysis of Lecanemab Treatment

In this section, we focus exclusively on the effects of Lecanemab by setting $u_D = 0$ and $u_A = 0$ in Eq. (15), similar to our approach in the previous subsection, we minimize the following objective function,

$$J(u_L) = \gamma F(T) + \int_0^T \left( j_1 N(t) + j_2 F(t) + \frac{1}{2}\alpha e^{-\beta t}\left(\frac{u_L(t)}{U_{max}^L}\right)^2 + c_L S_L(u_L)F(t) \right) dt.$$

and derive the adjoint equations for this case in Eq. (S14) given in **Supplementary Text E** For Lecanemab, $U_{max}^L$, the maximum clearance rate of amyloid-beta plaques induced by the drug, can be calculated from clinical data given in (36) as $U_{max}^L = 0.16\,\mu_6$. The side-effect function, $S_L(u_L)$, which is modeled as being proportional to the dosage, is expressed as:

$$S_L(u_L) = 0.0563\,\frac{u_L(t)}{U_{max}^L}.$$

For detailed calculations and parameter estimation for Lecanemab, refer to **Supplementary Text F**. Now, setting $\frac{\partial H}{\partial u_L} = 0$, on the interior of the control set and using the bounds, the characterization of the optimal control becomes:

$$u_{new}^L(t) = \min\left( U_{max}^L, \max\left( 0, (U_{max}^L)^2\,\frac{\lambda_P[P] + \lambda_F[F] - c_L\,0.0563\,\frac{F(t)}{U_{max}^L}}{\alpha e^{-\beta t}} \right) \right).$$

The resulting model is then solved numerically using the forward-backward sweep algorithm with boundary conditions. In Table 3, the row corresponding to treatment $L$ reflects the percentage reduction in fibril concentration, which is achieved exclusively through the control $u_L$, without any influence from $u_D$ or $u_A$.

## Analysis of Aducanumab treatment

In this section, we analyze the effects of Aducanumab by setting $u_D = 0$ and $u_L = 0$ in Eq. (15). Following the approach outlined in the previous subsections, we define the objective function for this treatment as:

$$J(u_A) = \gamma F(T) + \int_0^T \left( j_1 N(t) + j_2 F(t) + \frac{1}{2}\alpha e^{-\beta t}\left(\frac{u_A(t)}{U_{max}^A}\right)^2 + c_A S_A(u_A)F(t) \right) dt.$$

We then derive the adjoint equations for this case, as detailed in Eq. (S15) in **Supplementary Text E**.

Aducanumab's maximum dosage ($U_{\max}^A$) reflects the highest clearance rate of amyloid-beta plaques facilitated by the drug. Based on clinical data (Budd Haeberlein et al. 2022), we estimate $U_{\max}^A = 0.18\, \mu_6$. The associated side-effect function $S_A(u_A)$, which models ARIA-E incidence as proportional to the administered dosage, is expressed as:

$$S_A(u_A) = 0.301\, \frac{u_A(t)}{U_{\max}^A}.$$

For a detailed derivation and parameter estimation for Aducanumab, refer to **Supplementary Text F**. By setting $\frac{\partial H}{\partial u_A} = 0$ on the interior of the control set, we derive the optimal control $u_A(t)$ on that set. Taking the bounds into account, $0 \leq u_A(t) \leq U_{\max}^A$, the characterization of the optimal control becomes:

$$u_{\text{new}}^A(t) = \min\left( U_{\max}^A, \max\left( 0, (U_{\max}^A)^2\, \frac{\lambda_P[P] + \lambda_N[N] + \lambda_F[F] - c_A\, 0.301\, \frac{F(t)}{U_{\max}^A}}{\alpha e^{-\beta t}} \right) \right).$$

The positivity of the control function is maintained by selecting $c_A = 0.1$ in this case. The optimal control problem for Aducanumab is then solved, and the percentage reduction in fibril concentration achieved solely by this drug ($u_A \neq 0$ with $u_D = u_L = 0$) is reported in the entry labeled as treatment $A$ in Table 3.

**Analysis of ALL treatments**

To analyze the effectiveness of different treatments, we derive the adjoint equations Eq. (S16) by using Eq. (16) as the cost function and Eq. (15) as the state equations. Then We used these equations to solve the optimal control problem for various treatment scenarios by specifying which controls ($u_D, u_L, u_A$) are active. The results are summarized in Table 3, which shows the percentage reduction in fibril concentration ($F$) for different initial values of $M(0)$ under various treatment combinations.

In Table 3, the "All Drugs" scenario considers all three treatments—Donanemab ($u_D$), Lecanemab ($u_L$), and Aducanemab ($u_A$)—simultaneously. Here, all controls are active ($u_D, u_L, u_A \neq 0$), and the reductions in $F$ reflect the outcome of solving the full system, including all terms in the adjoint equations, cost function, and state equations.

The "L and A" scenario evaluates fibril reduction when Donanemab is excluded ($u_D = 0$), leaving only Lecanemab and Aducanemab ($u_L$ and $u_A$) active. Similarly, the "A and D" scenario examines the effect of combining Aducanemab and Donanemab ($u_A$ and $u_D$), with Lecanemab set to zero ($u_L = 0$). Finally, the "L and D" scenario explores the combination of Lecanemab and Donanemab ($u_L$ and $u_D$), excluding Aducanemab ($u_A = 0$).

Figures 5 and S8 illustrate the optimal control functions $u(t)$ for two treatment scenarios, each with different initial monomer concentrations $M(0)$:
1. **Single Treatment**: A scenario where only one drug is administered, meaning the corresponding control $u_i$ is nonzero while the other controls are set to zero.

2. **Combined Treatment**: A scenario where all three drugs ($u_D$, $u_L$, and $u_A$) are administered together, with all controls active.

We also compute the 95% confidence intervals for fibril concentrations $F(t)$ under ten different initial monomer concentrations $M(0)$, employing the same perturbation method for non-identifiable parameters outlined in Section **Uncertainty Quantification**, as depicted in Figure 6.

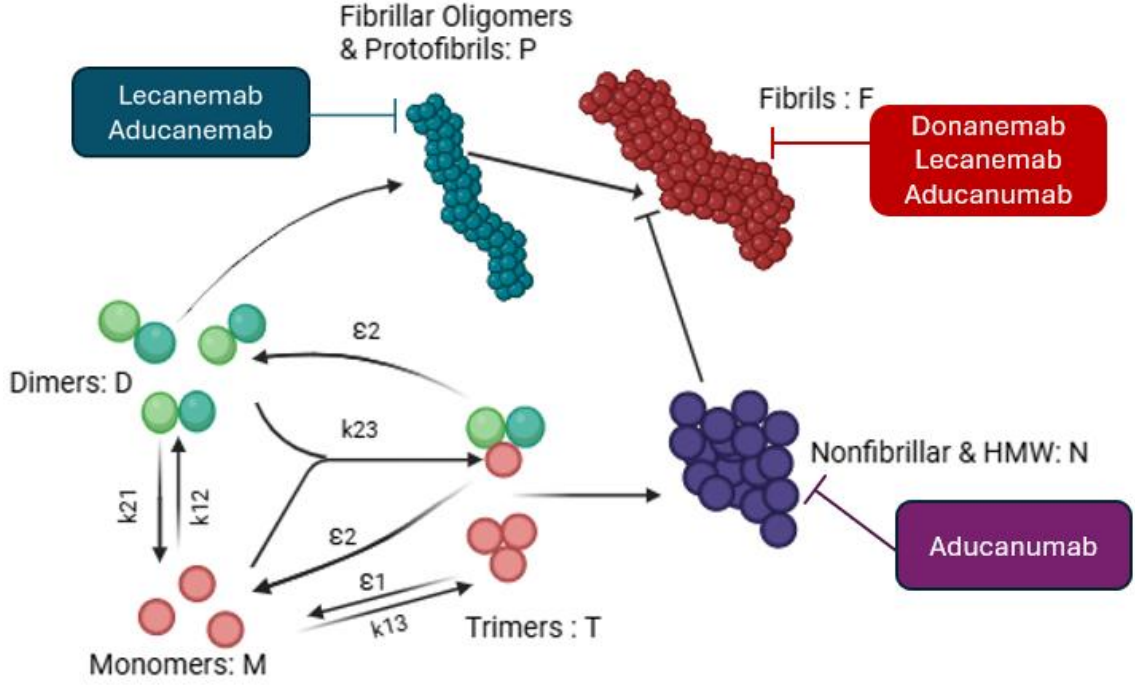

**Fig. 1:** The figure illustrates the aggregation of A$\beta$ monomers into oligomers and fibrils, with inhibitory arrows (blunt-ended lines) representing the action of factors that impede the conversion of oligomers to fibrils. The labeled boxes next to each aggregation species indicate the corresponding antibodies involved in their inhibition or clearance. Donanemab, Lecanemab, and Aducanumab target fibrils ($F$), while Lecanemab and Aducanumab also inhibit fibrillar oligomers and protofibrils ($P$). Additionally, Aducanumab targets nonfibrillar aggregates and high-molecular-weight species ($N$).

**Table 1:** Estimated parameter values and ranges for the mathematical model shown in Eqs. (2)–(8). We take $\delta = 0.1$ to define the variability for the corresponding parameters in our study.

| Parameters | Description | Value | Unit | Reference |
|---|---|---|---|---|
| $k_{12}$ | $2M \rightarrow D$ | $800 - 1180$ | $M^{-1}s^{-1}$ | (24) |
| $k_{21}$ | $D \rightarrow 2M$ | $8 \times 10^{-10} - 1.18 \times 10^{-9}$ | $s^{-1}$ | (24) |
| $k_{13}$ | $3M \rightarrow T$ | $10^9(1 \pm \delta)$ | $M^{-2}s^{-1}$ | Estimated |
| $\epsilon_1$ | $T \rightarrow 3M$ | $10^{-13} - 10^{-12}$ | $s^{-1}$ | (8) |
| $k_{23}$ | $M + D \rightarrow T$ | $33 - 43$ | $M^{-1}s^{-1}$ | (24) |

| Parameters | Description | Value | Unit | Reference |
|---|---|---|---|---|
| $\epsilon_2$ | $T \to M + D$ | $10^{-11} - 10^{-9}$ | $s^{-1}$ | (8) |
| $\mu_1$ | monomers clearance | $10^{-5} - 10^{-3}$ | $s^{-1}$ | (25) |
| $\mu_2$ | dimer clearance | $10^{-8}(1 \pm \delta)$ | $s^{-1}$ | Estimated |
| $\mu_3$ | trimer clearance | $10^{-8}(1 \pm \delta)$ | $s^{-1}$ | Estimated |
| $\mu_4$ | $P$ clearance | $10^{-4} - 10^{-3}$ | $s^{-1}$ | (25) |
| $\mu_5$ | $N$ clearance | $10^{-4}(1 \pm \delta)$ | $s^{-1}$ | (25) |
| $\mu_6$ | Fibrils clearance | $10^{-5} - 10^{-4}$ | $s^{-1}$ | (25) |
| $\hat{\mu}_1$ | Monomers birth | $10^{-14} - 10^{-12}$ | $Ms^{-1}$ | (27) |
| $\lambda_1$ | $\tilde{D} \to P$ | $10^5 - 10^7$ | $M^{-1}s^{-1}$ | Estimated |
| $\lambda_2$ | $\tilde{T} \to N$ | $10^3 - 10^4$ | $M^{-1}s^{-1}$ | Estimated |
| $\lambda_3$ | $P + N \to F$ | $0 - 1$ | $M^{-1}s^{-1}$ | Estimated |
| $K_1$ | dimer saturation | $10^{-7} - 4 \times 10^{-6}$ | $M$ | Assumed |
| $K_2$ | trimer saturation | $10^{-6} - 4 \times 10^{-6}$ | $M$ | Assumed |
| $K_3$ | N saturation | $10^{-3} - 5 \times 10^{-3}$ | $M$ | Assumed |

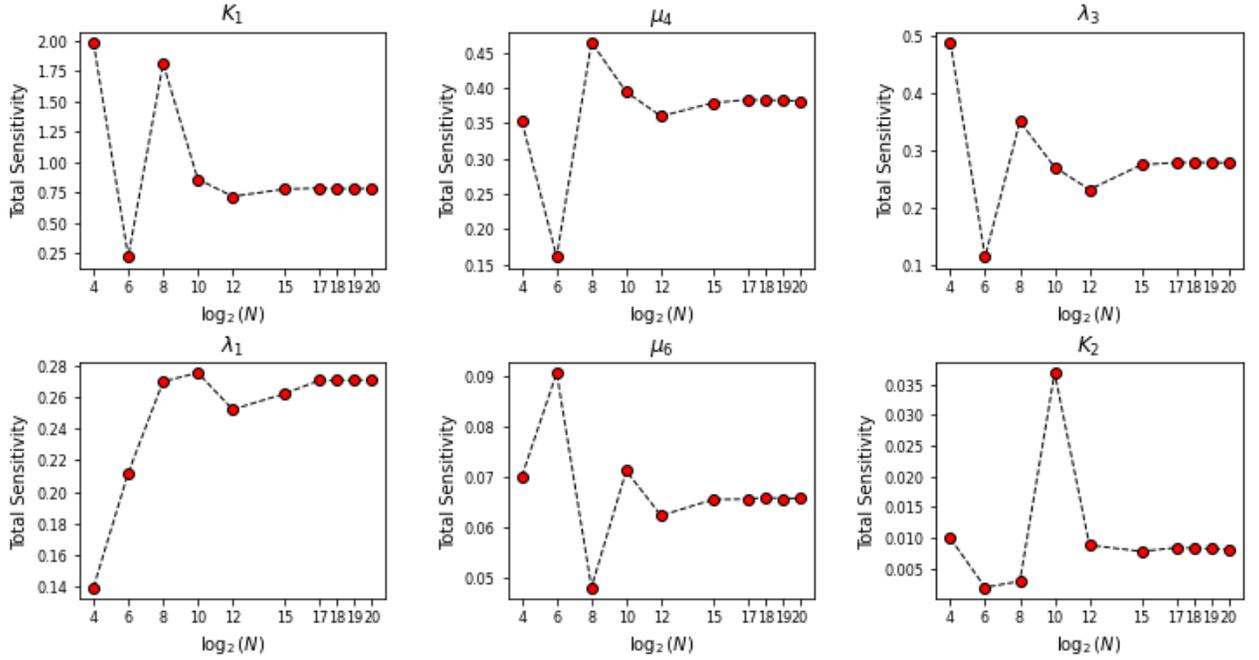

**Fig. 2:** Total sensitivity indices of $F(T)$ for the six most sensitive parameters with different sample sizes $N$ and the initial condition $M(0) = 5(\mu M)$.

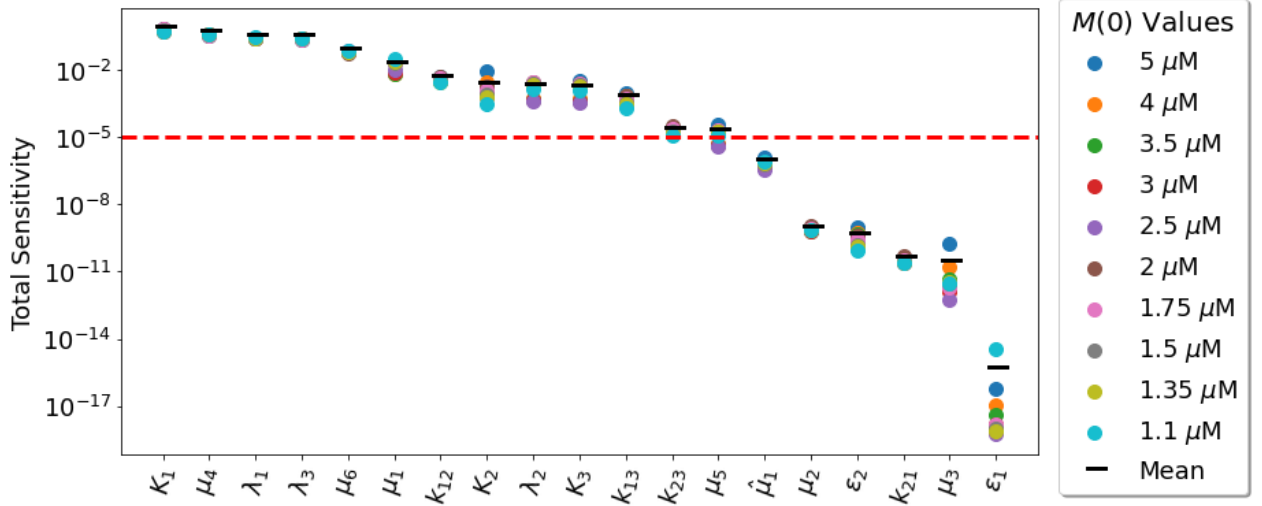

**Fig. 3:** Sobol total sensitivity indices for each parameter with 10 different initial conditions, showing total sensitivity indices for each initial condition (dot plot) and the average value (a dashed line). The sensitivity threshold $10^{-5}$ is also used to divide sensitive and non-sensitive parameters.

**Table 2:** Parameter ranges for different initial monomer concentrations.

| Initial Conditions | Parameter Ranges |
|---|---|
| $M(0) > 2\mu M$ | $K_1 = [10^{-7}, 4 \times 10^{-6}]\ K_2 = [10^{-6}, 4 \times 10^{-6}]\ \lambda_2 = [10^3, 10^4]$ |
| $M(0) \leq 2\mu M$ | $K_1 = [10^{-7}, 2 \times 10^{-6}]\ K_2 = [10^{-6}, 2 \times 10^{-6}]\ \lambda_2 = [10^4, 10^5]$ |

**Table 3:** Percentage reduction in fibril concentration, computed using the formula $\frac{F(T)-F_{u(T)}}{F(T)} \times 100$ for various initial monomer concentrations and different treatment plans. Here, $F(T)$ denotes the fibril concentration at the final time without treatment, while $F_{u(T)}$ represents the fibril concentration at the final time under the corresponding treatment.

|  | 5μM | 4μM | 3.5μM | 3μM | 2.5μM | 2μM | 1.75μM | 1.5μM | 1.35μM | 1.1μM |
|---|---|---|---|---|---|---|---|---|---|---|
| D | 85.7 | 85.8 | 85.9 | 86.1 | 85.9 | 86.0 | 86.0 | 85.9 | 85.9 | 85.8 |
| L | 15.6 | 17.6 | 17.9 | 20.7 | 20.8 | 16.0 | 15.5 | 16.1 | 20.6 | 24.9 |
| A | 17.0 | 19.1 | 19.2 | 22.4 | 24.5 | 14.4 | 12.1 | 17.4 | 21.9 | 27.0 |
| A and L | 28.6 | 31.9 | 32.1 | 36.8 | 41.1 | 26.5 | 24.0 | 29.2 | 36.2 | 43.2 |
| A and D | 86.5 | 86.9 | 86.9 | 87.4 | 87.8 | 86.1 | 86.3 | 86.6 | 87.3 | 88.2 |
| L and D | 86.4 | 86.6 | 86.9 | 87.4 | 88.1 | 86.6 | 86.6 | 86.7 | 87.3 | 88.0 |
| All drugs | 87.1 | 87.7 | 87.8 | 88.6 | 89.4 | 86.8 | 86.9 | 87.3 | 88.5 | 89.1 |

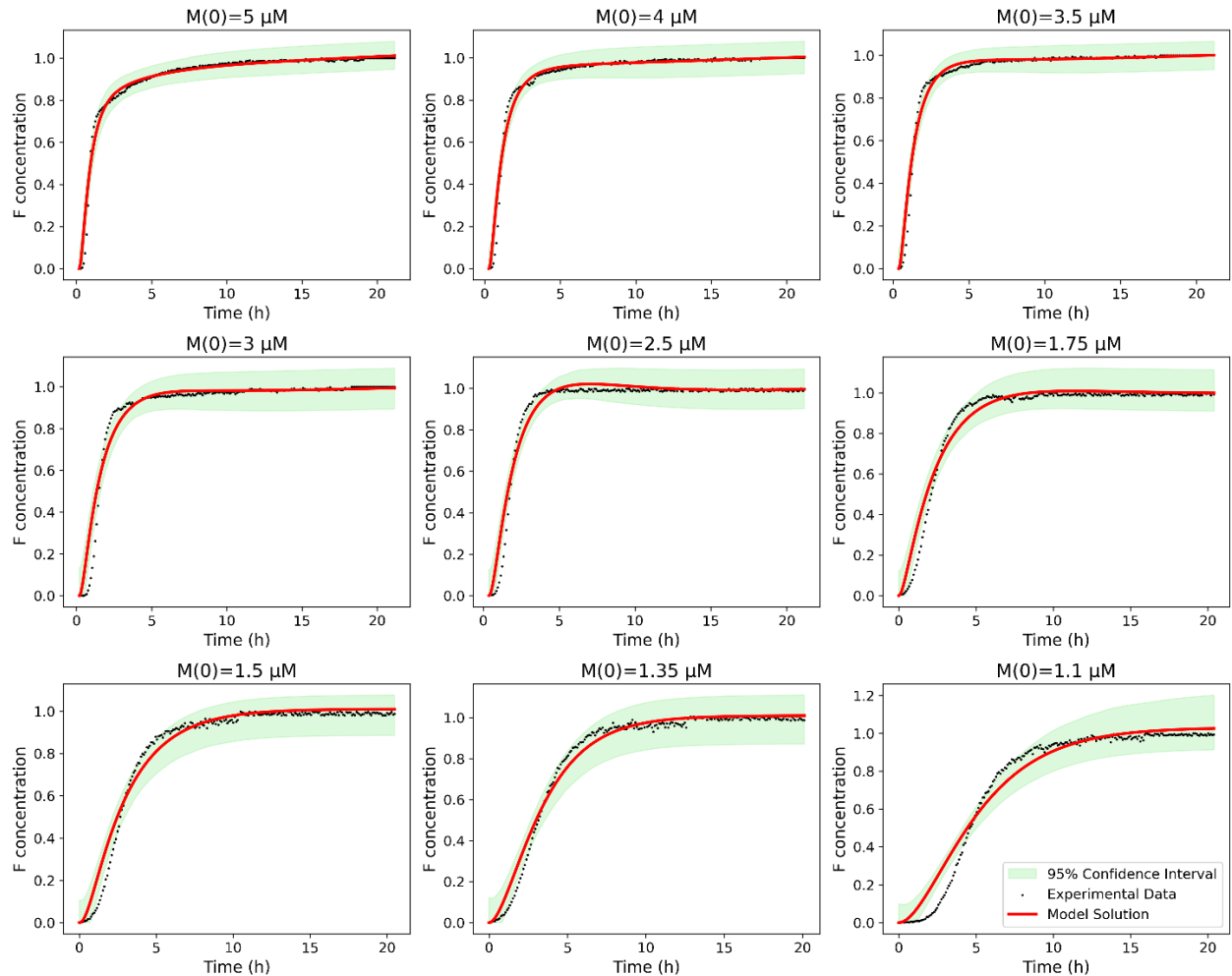

**Fig. 4:** Confidence intervals for nine different initial concentrations of $M(0)$, accounting for uncertainties introduced by non-identifiable parameters. Red curves represent simulated solutions using the optimal parameter vector $\tilde{\theta}$, while black dots correspond to experimental data. The 95% confidence intervals (green-shaded regions) reflect how uncertainties in non-identifiable parameters influence model predictions across different initial conditions.

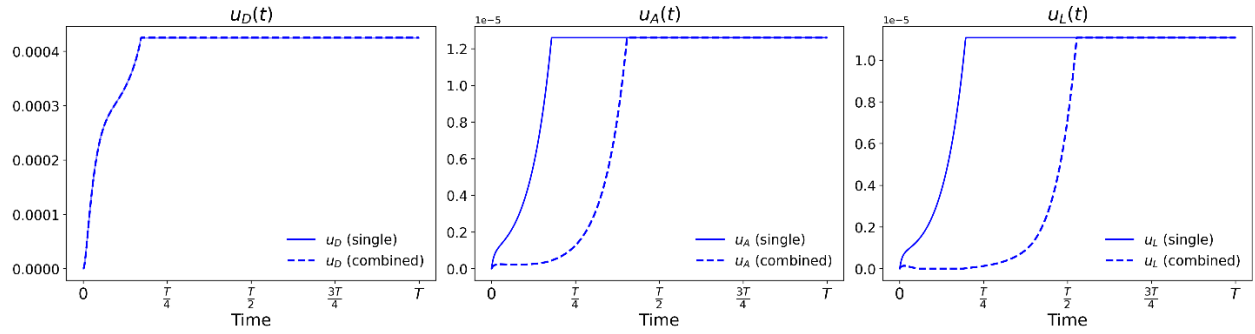

**Fig. 5:** Optimal control function values initial condition of $M(0) = 5\mu M$, comparing single-drug treatments (Left: Donanemab, $u_D$; Center: Aducanumab, $u_A$); Right: Lecanemab, $u_L$ with those obtained from combined therapy scenarios

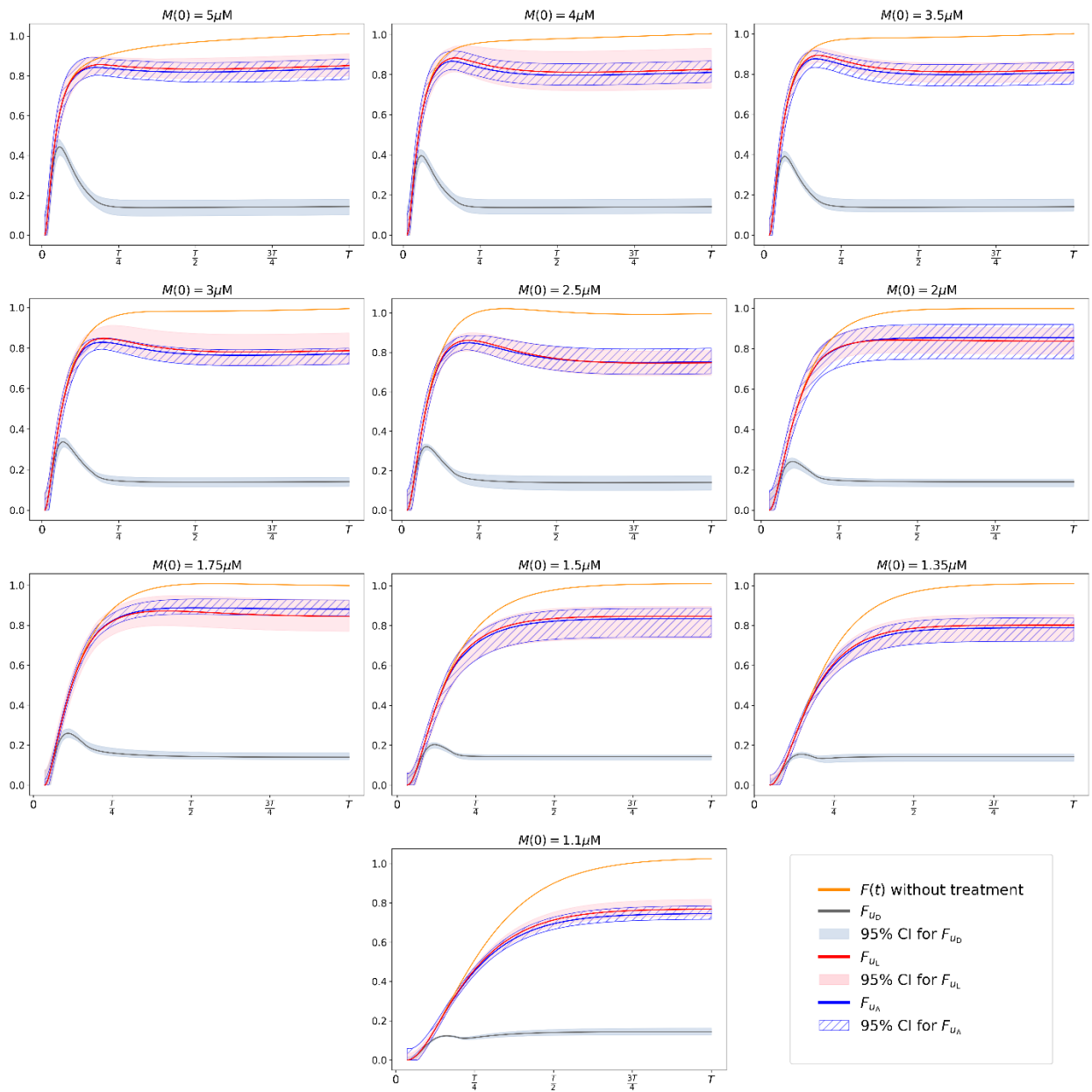

**Fig. 6:** 95% confidence intervals for $F$ after single-drug treatments with $u_D$, $u_L$, and $u_A$ (shaded in green, pink, and blue, respectively). The orange curve represents the $F$ concentration before treatment, plotted for ten different initial concentrations of $M(0)$.

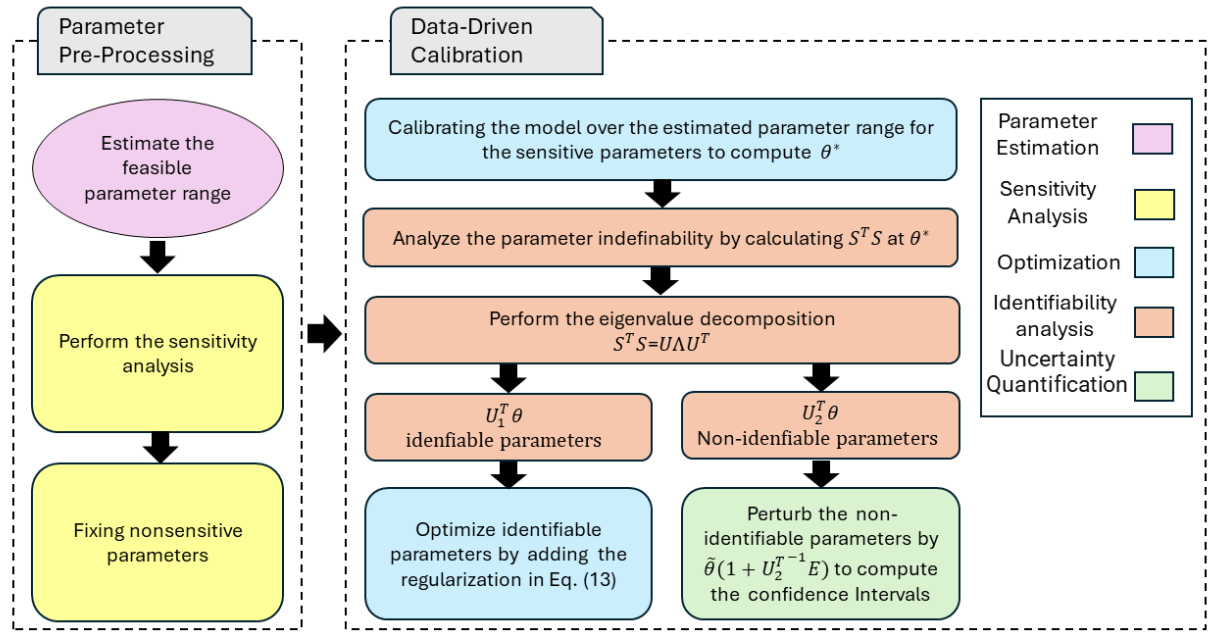

**Fig. 7:** The flowchart of the data-driven modeling approach. Parameter preprocessing produces the feasible parameter range summarized in Table 1, and the sensitivity analysis identifies the non-sensitive parameters, as shown in Table S3. Our data-driven approach calibrates the sensitive parameters using experimental data. We also perform parameter identifiability analysis to categorize these sensitive parameters into identifiable parameters, which can be calibrated using experimental data, and non-identifiable parameters, which contribute to the model's uncertainty but cannot be identified using the experimental data

# Supplementary Text

A: The Law of Mass Action

We consider a single reversible chemical reaction equation below:

$$\alpha A + \beta B \underset{k_r}{\overset{k_f}{\rightleftharpoons}} \gamma C. \tag{S1}$$

Here, $\alpha$, $\beta$, and $\gamma$ are stoichiometric coefficients. We will consider $[A_0]$, $[B_0]$, and $[C_0]$ as the initial (equilibrium) concentrations of $A$, $B$, and $C$ respectively. We can write the following equations to represent the concentrations of $A$, $B$, and $C$ at time $t$:

$$\begin{cases} [A] = [A_0] - \alpha n \\ [B] = [B_0] - \beta n \\ [C] = [C_0] + \gamma n \end{cases} \tag{S2}$$

where $n$ shows the number of times that reaction (S1) happens. Now if we eliminate $n$ in Eqs. (S2), we get

$$\gamma[A] + \alpha[C] = \text{constant} \quad \gamma[B] + \beta[C] = \text{constant}, \tag{S3}$$

which represents the conservation of elements in the reaction system. Now, let's consider the rates of change of the concentrations of species $A$, $B$, and $C$ to time:

$$\frac{d[A]}{dt} = \alpha r, \quad \frac{d[B]}{dt} = \beta r, \quad \frac{d[C]}{dt} = -\gamma r,$$

where $r$ is the rate of the reaction. Let's define $R_t = r$. By solving these equations, we obtain the following expressions for concentration:

$$[A] = \alpha R(t) + [A_0], \quad [B] = \beta R(t) + [B_0], \quad [C] = -\gamma R(t) + [C_0]. \tag{S4}$$

It has been established that a Lyapunov functional, also known as the free energy of the system (38), exists for the reaction (S1) governed by the law of mass action. The free energy can be expressed in various equivalent forms. In the case of a reaction kinetic system that satisfies the detailed balance condition, the Lyapunov function of the system is given by (39):

$$F([A],[B],[C]) = [A]\left(\ln\left(\frac{[A]}{[A_0]}\right) - 1\right) + [B]\left(\ln\left(\frac{[B]}{[B_0]}\right) - 1\right) + [C]\left(\ln\left(\frac{[C]}{[C_0]}\right) - 1\right),$$

To provide a thermodynamic perspective, the Lyapunov function can be reformulated as (39):

$$\begin{aligned} F([A],[B],[C],U_A,U_B,U_C) &= [A](\ln[A] - 1) + [B](\ln[B] - 1) \\ &\quad + [C](\ln[C] - 1) + [A]U_A + [B]U_B + [C]U_C, \end{aligned} \tag{S5}$$

where the first three terms correspond to the entropy, and the constant $U_{()}$ is the internal energy per mole associated with each species. The choice of $U_{()}$ determines the equilibrium of the system, i.e.,

$$\alpha(\ln[A_0] + U_A) + \beta(\ln[B_0] + U_B) - \gamma(\ln[C_0] + U_C) = 0,$$

which indicates that

$$\ln\left(\frac{[A_0]^\alpha [B_0]^\beta}{[C_0]^\gamma}\right) = \gamma U_C - \alpha U_A - \beta U_B := \Delta U.$$

Here $\Delta U$ is the difference of internal energy between the state $\{\alpha A, \beta B\}$ and the state $\{\gamma C\}$. Then the equilibrium constant $K_{eq}$ is defined as

$$K_{eq} = \left(\frac{[A_0]^\alpha [B_0]^\beta}{[C_0]^\gamma}\right) = e^{\Delta U}. \tag{S6}$$

By using the reaction trajectory, we can reformulate the free energy $F([A],[B],[C],U_A,U_B,U_C)$ defined in Eq. (S5) in terms of $R(t)$. By using the energy-dissipation law, we have

$$\frac{dF}{dt}(R, U_A, U_B, U_C) = -D(R, R_t),$$

where $D(R, R_t)$ is the dissipation of the system. The specific choice of $D(R, R_t)$ determines the reaction kinetics, which can differ from quadratic behavior in $R_t$ due to chemical reactions being typically far from thermodynamic equilibrium (38). The law of mass action can be derived from the energy-dissipation principle by selecting:

$$D(R, R_t) = R_t \ln\left(\frac{R}{k_r[C]^\gamma} + 1\right).$$

Consequently, we find that

$$\frac{\delta F}{\delta R}(R) = -\ln\left(\frac{R}{k_r[C]^\gamma} + 1\right),$$

On the other hand, we have

$$-\frac{\delta F}{\delta R}(R) = \ln\left(\frac{[A]^\alpha [B]^\beta}{[C]^\gamma}\right) + \alpha U_A + \beta U_B - \gamma U_C,$$

which implies

$$\ln\left(\frac{[A]^\alpha [B]^\beta}{[C]^\gamma}\right) - \Delta U = \ln\left(\frac{R}{k_r[C]^\gamma} + 1\right)$$

Finally, we derive the following equation

$$R_t = k_r[C]^\gamma \left(\frac{[A]^\alpha [B]^\beta}{K_{eq}[C]^\gamma} - 1\right) = k_f[A]^\alpha [B]^\beta - k_r[C]^\gamma$$

This equation corresponds to the classical law of mass action. The last equality is obtained by utilizing the relation $K_{eq} = e^{\Delta U} = \frac{k_r}{k_f}$.

## B: Sobol Sensitivity Analysis

Any model can be viewed as a system $X = f(\boldsymbol{\theta})$, where $\boldsymbol{\theta}$ is a vector of $n_p$ uncertain model inputs $[\theta_1, \theta_2, \ldots, \theta_{n_p}]$ in $\mathbb{R}^{n_p}$ and $X$ is the model output. In complex models involving numerous inputs, understanding how variations in these inputs influence the output is crucial. Sobol sensitivity analysis offers a systematic and comprehensive method to assess this influence.

Sobol sensitivity analysis, developed by Ilya Sobol in the 1990s, is a global sensitivity analysis technique. It quantifies the contribution of each input variable and their interactions to the model output variance. This global approach evaluates the importance of each input across the entire parameter space, unlike local sensitivity analyses which focus on variations around a specific point in the input space.

**First-Order Sensitivity Analysis:**
The first-order sensitivity index quantifies the influence of each parameter within the vector $\boldsymbol{\theta}$ on the variance of the model's output when considering each parameter in isolation. Its calculation is as follows:

$$S_i(X) = \frac{\text{Var}_{\theta_i}\left[E_{\theta_{-i}}(X|\theta_i)\right]}{\text{Var}(X)} \tag{S7}$$

This adjusted expression represents the first-order sensitivity index for parameter $\theta_i$ more explicitly, where:

- $\text{Var}_{\theta_i}\left[E_{\theta_{-i}}(X|\theta_i)\right]$ represents the variance of the conditional expectation of $X$ with respect to all parameters except $\theta_i$, considering $\theta_i$ as the variable of interest.

- $\text{Var}(X)$ represents the variance of the model's output $X$.

**Second-Order Sensitivity Analysis:**

The second-order sensitivity, concerning parameters $\theta_i$ and $\theta_j$ within the vector $\boldsymbol{\theta}$, captures the cumulative impact of their first-order effects and their interaction at the second order. Its calculation is as follows:

$$S_{(i,j)}(X) = S_i(X) + S_j(X) + \frac{\text{Var}_{(\theta_i, \theta_j)}\left[E_{\theta_{-i,j}}(X|\theta_i, \theta_j)\right]}{\text{Var}(X)} \tag{S8}$$

This analysis permits us to explore how pairs of parameters jointly affect the model output.

**Total-Order Sensitivity Analysis:**
The total-order sensitivity index gauges the overall influence of a parameter, encompassing its first order and higher-order effects:

$$S_{T,i}(X) = 1 - \frac{\text{Var}_{-\theta_i}\left[E_{\theta_i}(X|-\theta_i)\right]}{\text{Var}(X)} \tag{S9}$$

The absolute value of the sensitivity quantifies the extent of the impact of a parameter on the model output. Parameters with sensitivity values closer to 0 influence the model's overall output less.

## C: Parameter Identifiability

For the data-driven modeling approach, a loss function $L(\theta)$ is defined as follows:

$$L(\theta) = \sum_{j=1}^{N} \left(y(t_j, \theta) - y_j\right)^2,$$

where $y(t_j, \theta)$ denotes the system output at time $t_j$ with parameters $\theta$, and $y_j$ is the corresponding experimental data without error.

$$\theta^* = argminL(\theta).$$

Practical identifiability is the ability to estimate model parameters uniquely and accurately using the available experimental data. This involves examining how sensitive the model output is to changes in parameter values and ensuring that the parameters can be distinguished based on their influence on the model predictions. Practical identifiability analysis helps to determine model parameters with a reasonable level of confidence.

One approach to practical identifiability analysis is sensitivity-based identifiability analysis. This method examines identifiability at specific points. It uses the sensitivity of measurable system responses to parameter values to assess the identifiability of unknown parameters. More specifically, assume that the locations and the number of time points at which the system responses or state variables will be measured are given, denoted by $t_1 \leq t_2 \leq \cdots \leq t_N$. The sensitivity coefficient at each time point $t_k$ ($k = 1,2,\ldots,N$) for a given nominal parameter vector $\theta^*$ is defined as:

$$s_{ij}(t_k) = \frac{\partial y_i(t_k, \theta^*)}{\partial \theta_j},$$

where $y_i$ ($i = 1,2,\ldots,d$) denotes the $i$-th component of $y$ ($y \in \mathbb{R}^d$) and $\theta_j$ ($j = 1,2,\ldots,q$) the $j$-th component of $\theta$ ($\theta \in \mathbb{R}^q$). Thus, the sensitivity matrix for all time points is defined as:

$$S_{dN \times q} = \begin{bmatrix} s_{11}(t_1) & \cdots & s_{1q}(t_1) \\ \vdots & \ddots & \vdots \\ s_{d1}(t_1) & \cdots & s_{dq}(t_1) \\ \vdots & \ddots & \vdots \\ s_{11}(t_N) & \cdots & s_{1q}(t_N) \\ \vdots & \ddots & \vdots \\ s_{d1}(t_N) & \cdots & s_{dq}(t_N) \end{bmatrix}.$$

Several identifiability analysis techniques have been developed based on this sensitivity matrix. Generally, the larger the sensitivity coefficients, the more significant the system responses or measurable state variables are with respect to changes in parameters. In that sense, a parameter is likely to be identifiable if the system output is highly sensitive to a small change in that parameter; otherwise, the parameter is likely to be unidentifiable. Additionally, if there is a strong correlation between any two parameters, those two parameters are very likely to be indistinguishable from each other. Such parameter dependence can also be evaluated by examining the dependence of the sensitivity matrix columns (40).

Our goal is to modify this approach to include the presence of noise in measurements, as errors are typically unavoidable in practical applications. Thus, we define the loss function with noise term, $L_\epsilon(\theta)$, as:

$$L_\epsilon(\theta) = \sum_{j=1}^{N} \left(y(t_j, \theta) - y_j - \epsilon\right)^2,$$

and proceed with the gradient calculation as follows:

$$\nabla L_\epsilon(\theta) = 2 \sum_{j=1}^{N} \left(y(t_j, \theta) - y_j - \epsilon\right) \nabla y(t_j, \theta).$$

Then, we employ a Taylor expansion for $y(t_j, \theta)$ and accordingly for $\nabla y(t_j, \theta)$, resulting in:

$$y(t_j, \theta) = y(t_j, \theta^*) + \nabla y(t_j, \theta^*)^T (\theta - \theta^*) + \frac{1}{2}(\theta - \theta^*)^T H\left(y(t_j, \theta^*)\right)(\theta - \theta^*) + O(\|\theta - \theta^*\|^3),$$

and

$$\nabla y(t_j, \theta) = \nabla y(t_j, \theta^*) + H\left(y(t_j, \theta^*)\right)(\theta - \theta^*) + O(\|\theta - \theta^*\|^2),$$

where $H\left(y(t_j, \theta^*)\right)$ denotes the Hessian matrix of $y$ with respect to $\theta$, evaluated at $\theta^*$. By inserting the Taylor expansions into $\nabla L_\epsilon(\theta)$ and considering only the linear components of $(\theta - \theta^*)$, We derive:

$$\nabla L_\epsilon(\theta) = \sum_{j=1}^{N} \left(\left(y(t_j, \theta^*) - y_j\right) \nabla y(t_j, \theta^*) + \nabla y(t_j, \theta^*) \nabla y(t_j, \theta^*)^T (\theta - \theta^*)\right.$$
$$\left. + \left(y(t_j, \theta^*) - y_j - \epsilon\right) H(\theta - \theta^*) - \epsilon \nabla y(t_j, \theta^*)\right) + O(\|\theta - \theta^*\|^2). \quad (S10)$$

Since the first term $\nabla L(\theta^*) = 0$, if the residual $|L(\theta^*)| \leq \epsilon$ we simplify $\nabla L_\epsilon(\theta)$ as follows:

$$\nabla L_\epsilon(\theta) = \sum_{j=1}^{N} \nabla y(t_j, \theta^*) \nabla y(t_j, \theta^*)^T (\theta - \theta^*) + O(\epsilon) + O(\|\theta - \theta^*\|^2).$$

Next, we introduce the sensitivity $s_k(t_j) = \frac{\partial y(t_j, \theta^*)}{\partial \theta_k}$ and the sensitivity matrix $S$, defined as follows

$$S = \begin{bmatrix} s_1(t_1) & s_2(t_1) & \cdots & s_q(t_1) \\ s_1(t_2) & s_2(t_2) & \cdots & s_q(t_2) \\ \vdots & \vdots & \ddots & \vdots \\ s_1(t_N) & s_2(t_N) & \cdots & s_q(t_N) \end{bmatrix}, \quad (S11)$$

thus, we have

$$\nabla L_\epsilon(\theta) = S^T S(\theta - \theta^*) + O(\epsilon) + O(\|\theta - \theta^*\|^2) = 0. \tag{S12}$$

Then the identifiability transfers to solve the above linear system for any given $\epsilon$, if $S^T S$ is full rank, $\theta - \theta^* = O(\epsilon)$ is identifiable. If $S^T S$ is singular, then there exists at least one nontrivial solution $\hat{\theta} \neq \theta^*$ such that the model parameters are not identifiable at $\theta^*$.

### D: Optimized parameter values

We applied ten sets of experimental data (normalized), each illustrating the concentration of fibrils from ten distinct initial monomer concentrations, denoted by $M(0)$. Utilizing the enhanced calibration method outlined in Equation 13, we determined the optimal parameter vector $\tilde{\theta}$ for our model under each of these initial conditions. Table S2 lists the values of the parameters in these optimal vectors.

### E: Adjoint Equations

This appendix provides a comprehensive derivation of the adjoint equations for the optimal control problem associated with treatments using Donanemab, Lecanemab, Aducanumab, and their combined therapy.

*General Optimal Control Framework*

The state dynamics of the system are given by:

$$\frac{dx}{dt} = f(x(t), u(t)), \quad x(0) = x_0,$$

with a cost functional:

$$J = \phi(x(T)) + \int_0^T L(x(t), u(t))\, dt.$$

The Hamiltonian is defined as:

$$H(x(t), u(t), \Lambda(t)) = L(x(t), u(t)) + \Lambda^T f(x(t), u(t)),$$

where $\Lambda(t)$ represents the adjoint variables.

Using Pontryagin's Maximum Principle, the adjoint equations are derived as follows:
State Dynamics:

$$\frac{dx}{dt} = \frac{\partial H}{\partial \Lambda}.$$

Adjoint Dynamics:

$$\frac{d\Lambda}{dt} = -\frac{\partial H}{\partial x}.$$

Optimality Condition:
$$\frac{\partial H}{\partial u} = 0.$$

Boundary conditions:
$$\Lambda(T) = \frac{\partial \phi}{\partial x}\Big|_{x(T)}.$$

*Adjoint Equations for Donanemab*

Setting $u_D \neq 0$ and $u_L = u_A = 0$, the Hamiltonian is:
$$H = j_1 N + j_2 F + \frac{1}{2}\alpha e^{-\beta t}\left(\frac{u_D}{U_{\max}^D}\right)^2 + c_D S_D(u_D)F + \Lambda^\mathsf{T} f(x, u_D).$$

Using Pontryagin's Maximum Principle, the adjoint equations are derived as:

$$\frac{d\lambda_M}{dt} = \lambda_M(2k_{12}[M] + 3k_{13}[M]^2 + k_{23}[D] + \mu_1) - \lambda_D(2k_{12}[M] - k_{23}[D]) - \lambda_T(3k_{13}[M]^2 + k_{23}[D]),$$

$$\frac{d\lambda_D}{dt} = \lambda_M(-k_{21} + k_{23}[M]) + \lambda_D(k_{21} + k_{23}[M] + \mu_2) - \lambda_T(k_{23}[M]) - \lambda_P(\lambda_1(K_1 - 2[D])),$$

$$\frac{d\lambda_T}{dt} = -\lambda_M(\epsilon_1 + \epsilon_2) - \lambda_D(\epsilon_2) + \lambda_T(\epsilon_1 + \epsilon_2 + \mu_3) - \lambda_N(\lambda_2(K_2 - 2[T])),$$

$$\frac{d\lambda_P}{dt} = \mu_4 \lambda_P - \frac{\lambda_3 \lambda_F}{1 + [N]/K_3},$$

$$\frac{d\lambda_N}{dt} = -j_1 + \mu_5 \lambda_N + \frac{\lambda_3 \lambda_F [P]/K_3}{(1 + [N]/K_3)^2},$$

$$\frac{d\lambda_F}{dt} = -j_2 + \mu_6 \lambda_F + u_D(t)\lambda_F - c_D S_D. \tag{S13}$$

with the following boundary conditions:
$$\lambda_M = \lambda_D = \lambda_T = \lambda_P = \lambda_N = 0, \quad \lambda_F = \gamma$$

*Adjoint Equations for Lecanemab*

For $u_L \neq 0$ and $u_D = u_A = 0$, the Hamiltonian adapts to include $u_L$. The adjoint equations are:

$$\frac{d\lambda_M}{dt} = \lambda_M(2k_{12}[M] + 3k_{13}[M]^2 + k_{23}[D] + \mu_1) - \lambda_D(2k_{12}[M] - k_{23}[D]) - \lambda_T(3k_{13}[M]^2 + k_{23}[D]),$$

$$\frac{d\lambda_D}{dt} = \lambda_M(-k_{21} + k_{23}[M]) + \lambda_D(k_{21} + k_{23}[M] + \mu_2) - \lambda_T(k_{23}[M]) - \lambda_P\big(\lambda_1(K_1 - 2[D])\big),$$

$$\frac{d\lambda_T}{dt} = -\lambda_M(\epsilon_1 + \epsilon_2) - \lambda_D(\epsilon_2) + \lambda_T(\epsilon_1 + \epsilon_2 + \mu_3) - \lambda_N\big(\lambda_2(K_2 - 2[T])\big),$$

$$\frac{d\lambda_P}{dt} = \big(\mu_4 + u_L(t)\big)\lambda_P - \frac{\lambda_3 \lambda_F}{1 + \frac{[N]}{K_3}},$$

$$\frac{d\lambda_N}{dt} = -j_1 + \mu_5 \lambda_N + \frac{\frac{\lambda_3 \lambda_F [P]}{K_3}}{\left(1 + \frac{[N]}{K_3}\right)^2},$$

$$\frac{d\lambda_F}{dt} = -j_2 + \big(\mu_6 + u_L(t)\big)\lambda_F - c_L S_L. \tag{S14}$$

*Adjoint Equations for Aducanumab*

For $u_A \neq 0$ and $u_D = u_L = 0$, the adjoint equations are obtained as:

$$\frac{d\lambda_M}{dt} = \lambda_M(2k_{12}[M] + 3k_{13}[M]^2 + k_{23}[D] + \mu_1) - \lambda_D(2k_{12}[M] - k_{23}[D]) - \lambda_T(3k_{13}[M]^2 + k_{23}[D]),$$

$$\frac{d\lambda_D}{dt} = \lambda_M(-k_{21} + k_{23}[M]) + \lambda_D(k_{21} + k_{23}[M] + \mu_2) - \lambda_T(k_{23}[M]) - \lambda_P\big(\lambda_1(K_1 - 2[D])\big),$$

$$\frac{d\lambda_T}{dt} = -\lambda_M(\epsilon_1 + \epsilon_2) - \lambda_D(\epsilon_2) + \lambda_T(\epsilon_1 + \epsilon_2 + \mu_3) - \lambda_N\big(\lambda_2(K_2 - 2[T])\big),$$

$$\frac{d\lambda_P}{dt} = \big(\mu_4 + u_A(t)\big)\lambda_P - \frac{\lambda_3 \lambda_F}{1 + [N]/K_3},$$

$$\frac{d\lambda_N}{dt} = -j_1 + \big(\mu_5 + u_A(t)\big)\lambda_N + \frac{\lambda_3 \lambda_F [P]/K_3}{(1 + [N]/K_3)^2},$$

$$\frac{d\lambda_F}{dt} = -j_2 + \big(\mu_6 + u_A(t)\big)\lambda_F - c_A S_A. \tag{S15}$$

*Adjoint Equations for Combined Treatment*

When $u_D, u_L, u_A \neq 0$, the adjoint equations aggregate contributions from all controls:

$$\frac{d\lambda_M}{dt} = \lambda_M(2k_{12}[M] + 3k_{13}[M]^2 + k_{23}[D] + \mu_1) - \lambda_D(2k_{12}[M] - k_{23}[D]) - \lambda_T(3k_{13}[M]^2 + k_{23}[D]),$$

$$\frac{d\lambda_D}{dt} = \lambda_M(-k_{21} + k_{23}[M]) + \lambda_D(k_{21} + k_{23}[M] + \mu_2) - \lambda_T(k_{23}[M]) - \lambda_P\big(\lambda_1(K_1 - 2[D])\big),$$

$$\frac{d\lambda_T}{dt} = -\lambda_M(\epsilon_1 + \epsilon_2) - \lambda_D(\epsilon_2) + \lambda_T(\epsilon_1 + \epsilon_2 + \mu_3) - \lambda_N\big(\lambda_2(K_2 - 2[T])\big),$$

$$\frac{d\lambda_P}{dt} = \big(\mu_4 + u_L(t) + u_A(t)\big)\lambda_P - \frac{\lambda_3 \lambda_F}{1 + [N]/K_3},$$

$$\frac{d\lambda_N}{dt} = -j_1 + \big(\mu_5 + u_A(t)\big)\lambda_N + \frac{\lambda_3 \lambda_F [P]/K_3}{(1 + [N]/K_3)^2},$$

$$\frac{d\lambda_F}{dt} = -j_2 + \big(\mu_6 + u_D(t) + u_L(t) + u_A(t)\big)\lambda_F - c_D S_D - c_L S_L - c_A S_A. \tag{S16}$$

**F: Maximum Dosages and Side Effects**

In this section, we use clinical data to develop a dose-dependent side effect function. To ensure consistency, we focus on studies that report amyloid plaque reduction at the end of treatment, allowing us to estimate the maximum usable dosage in our mathematical model based on each drug's clearance rate.

We specifically selected studies that not only align with FDA-approved package inserts but also include additional data on alternative treatment regimens, such as different dosage levels and their associated side effects. By incorporating a broader dataset, we aim to refine the side effect function and establish a more accurate estimation of the maximum dosage that can be realistically applied in our model.

*Maximum Dosage of Donanemab ($U_{max}^D$)*

To estimate the maximum dosage of Donanemab ($U_{max}^D$), we use clinical data reported in (35), which states that at 76 weeks, brain amyloid plaque levels decreased by 88.0 (from 102.4 to 14.4). The following equation describes the dynamics of amyloid plaques:

$$\frac{dA\beta}{dt} = \Lambda - \mu A\beta - U_{max}^D A\beta, \tag{S17}$$

where:

- $\Lambda$: Production of amyloid plaques from smaller species aggregating into plaques,
- $-\mu A\beta$: Natural clearance rate of $A\beta$,
- $-U_{max}^D A\beta$: Drug-induced clearance of $A\beta$.

At equilibrium:

$$A\beta_I = \frac{\Lambda}{\mu}, \quad A\beta_F = \frac{\Lambda}{\mu + U_{max}^D},$$

where $A\beta_I$ and $A\beta_F$ are the initial and final plaque levels, respectively. From clinical data:

$$\frac{A\beta_F}{A\beta_I} = \frac{14.4}{102.4} \approx 0.14.$$

Substituting this into the equilibrium equations:

$$\frac{\mu}{\mu + U^D_{\max}} = 0.14 \Rightarrow U^D_{\max} = \frac{86}{14}\mu.$$

We incorporate this relationship between $U^D_{\max}$ and the natural clearance rate into our model and consider $U^D_{\max} = \frac{43}{7}\mu_6$.

*Side Effects of Donanemab ($S_D(u_D)$)*

Clinical data from (35) reports Amyloid-Related Imaging Abnormalities-Edema (ARIA-E) in 24.0% of the Donanemab group and 1.9% of the placebo group. The probability of ARIA-E is modeled as proportional to the dosage $u_D(t)$:

$$S_D(u_D) = 0.221 \frac{u_D(t)}{U^D_{\max}}.$$

This linear relationship links the side effects to the dosage administered during treatment.

*Maximum Dosage of Lecanemab ($U^L_{max}$)*

In (36), the baseline amyloid PET SUVR values for treatment regimens of 2.5 mg/kg biweekly, 5 mg/kg monthly, 5 mg/kg biweekly, 10 mg/kg monthly, and 10 mg/kg biweekly are reported as 1.41, 1.42, 1.40, 1.42, and 1.37, respectively. The corresponding changes from baseline are $-0.094, -0.131, -0.197, -0.225,$ and $-0.306$.

Table S4 summarizes the baseline and final amyloid plaque sizes for these Lecanemab treatment regimens. The baseline values represent the initial amyloid plaque size before treatment, while the final values reflect the amyloid plaque levels after the treatment period.
Using Eq. (S17) and following a similar approach as for Donanemab, we derive a relationship between $U^L_{\max}$ and $\mu_6$ for each treatment group. The results are summarized in Table S5, including the reported data for ARIA-E adverse effects for each treatment group as provided in the reference paper.
To ensure a unique maximum dosage in our model and in the characteristic equation for $u_L$, we calculate the average of the maximum dosages reported in this table and set $U^L_{max} = 0.16\,\mu_6$.

*Estimation of Side Effects of Lecanemab ($S_L(u_L)$)*

To maintain consistency in the side effect model across all treatments, similar to $S_D(t)$, we use the ARIA-E data provided in Table S5 to derive a linear fit for $S_L$ as:

$$S_L(u_L) = 0.0563 \frac{u_L(t)}{U^L_{\max}}.$$

Additionally, we consider $c_L = 1$ and substitute these terms into the optimal control model.

*Maximum Dosage of Aducanumab ($U_{max}^A$)*

To estimate $U_{max}^A$ for Aducanumab, we refer to the results reported in a clinical study (37). Participants were grouped into two categories: Emerge and Engage. The study included two dosage levels:

- Low Dose Group: Participants received either 3 mg/kg or 6 mg/kg every 4 weeks over 76 weeks, with an ARIA-E incidence of 26%.
- High Dose Group: Participants received 10 mg/kg every 4 weeks over 76 weeks, with an ARIA-E incidence of 35%.

We summarize the data for amyloid PET SUVR values and ARIA-E incidence in the Emerge and Engage groups in Table S6.

The table captures essential data for both dosage levels across the Emerge and Engage groups. Considering the mean values between these groups, we calculate $U_{max}^A$ based on $\mu_6$, following a similar approach as in previous cases. Table S7 reports the results:

To ensure a unique maximum dosage in our model and the characteristic equation for $u_A$, we calculate the average of the maximum dosages reported in this table and set $U_{max}^A = 0.18\,\mu_6$.

*Modeling Side Effects of Aducanumab ($S_A(u_A)$)*

For Aducanumab, we use the ARIA-E data provided in Table S7. The side-effect function is defined as:

$$S_A(u_A) = 0.301\,\frac{u_A(t)}{U_{max}^A}.$$

Fig. S8 shows the optimal control function values for different initial conditions $M(0)$. The control functions for single-drug treatments (Donanemab, $u_D$; Aducanumab, $u_A$; Lecanemab, $u_L$) are compared with those from combined therapy scenarios. For each initial condition, the solid and dashed lines corresponding to the single-drug and combined therapy are plotted in the same color.

**Algorithm 1:** Forward-Backward Sweep Algorithm with Boundary Conditions

---

1: **Initialize** the control $u_D(t)$ as a zero function.
2: **while** the relative error is not sufficiently small **do**
3:    Solve the state equations forward in time using the current control $u_D(t)$ to compute the state variables.
4:    Solve the adjoint equations backward in time using the current states and control $u_D(t)$ to compute the adjoint variables.
5:    Update the control function $u_D(t)$ using the optimal control characterization:

$$u_D(t) = \min\left(U_{max}^D, \max\left(0, (U_{max}^D)^2\,\frac{\lambda_F[F] - \frac{0.221\,c_D[F]}{U_{max}^D}}{\alpha_D e^{\beta_D t}}\right)\right).$$

6:    Compute the relative error of the control and check for convergence.
7: **end while**

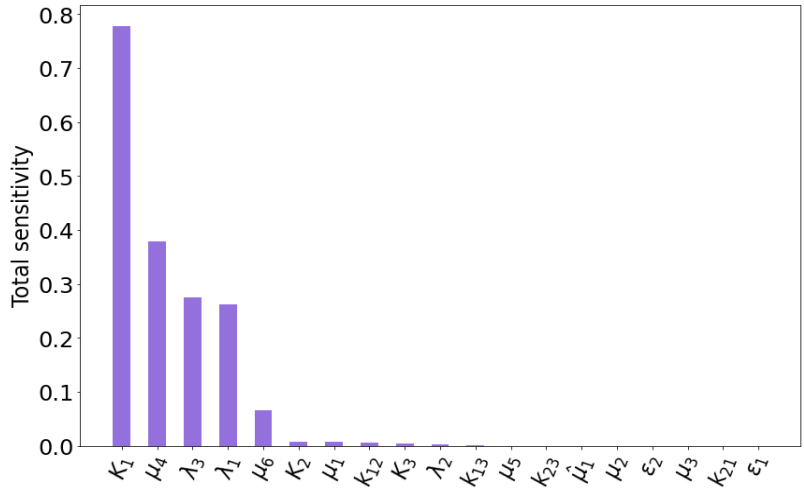

**Fig. S1**: Total sensitivity indices of $F(T)$ for different parameters for sample points $N = 2^{15}$. Here we choose the initial condition $M(0) = 5(\mu M)$.

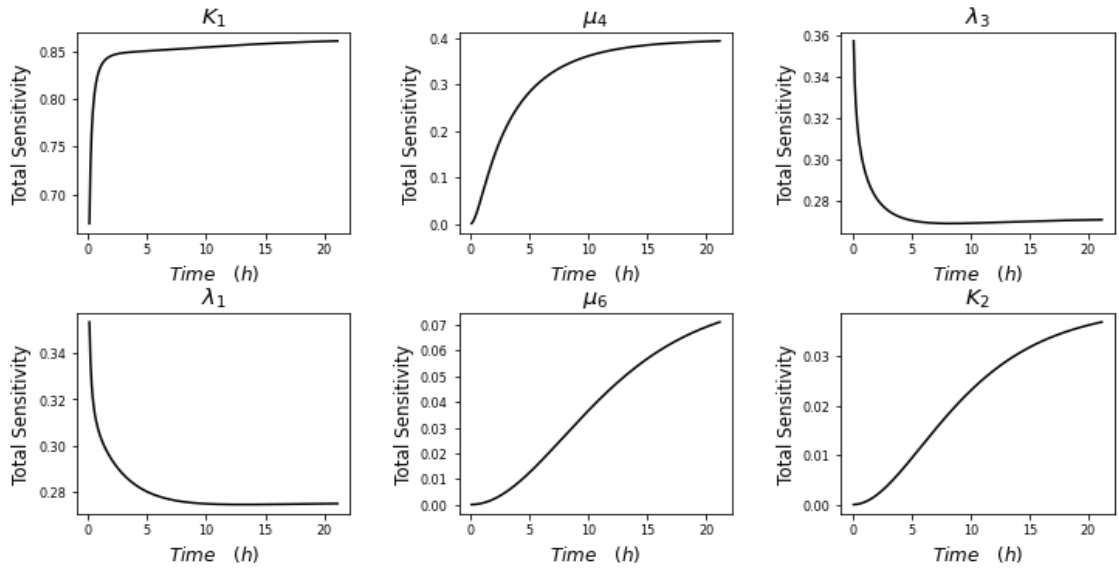

**Fig. S2**: Total sensitivity indices of $F(t)$ for the six most sensitive parameters at different times for $M(0) = 5\mu M$ and with $N = 2^{15}$.

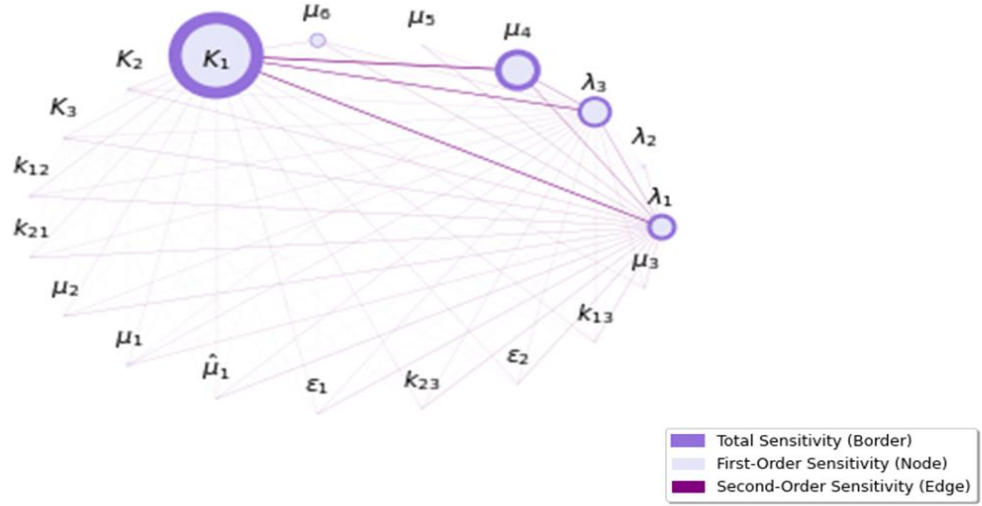

**Fig. S3**: Network representation of parameter sensitivities for $N = 2^{15}$, showing first, second, and total order effects on $F(T)$ with $M(0) = 5\mu M$. Node sizes represent the first-order Sobol sensitivity indices ($S_i$) for each parameter, while node borders highlight each parameter's total sensitivity ($S_{T,i}$), indicating their direct effect on $F(T)$. The thickness of the edges between nodes reflects the second-order sensitivity indices ($S_{(i,j)}$), indicating interaction effects between parameter pairs.

**Table S1**: Total Sobol sensitivity indices (ST) of $F(T)$ and their corresponding confidence intervals for a sample size of $N = 2^{20}$ across various parameters the initial condition $M(0) = 5(\mu M)$. The confidence intervals quantify the uncertainty in the estimation of the Sobol indices due to the sampling process in the sensitivity analysis.

| Parameter | $\lambda_1$ | $\lambda_2$ | $\lambda_3$ | $\mu_4$ | $\mu_5$ | $\mu_6$ | $K_1$ | $K_2$ | $K_3$ | $k_{12}$ |
|---|---|---|---|---|---|---|---|---|---|---|
| $S_T$ index | 2.71e-01 | 3.06e-03 | 2.78e-01 | 3.81e-01 | 7.07e-05 | 6.57e-02 | 7.81e-01 | 8.00e-03 | 3.26e-03 | 5.54e-03 |
| Conf Int | 4.93e-03 | 3.58e-04 | 4.31e-03 | 5.62e-03 | 7.12e-06 | 1.12e-03 | 1.33e-02 | 4.44e-04 | 2.22e-04 | 1.38e-04 |

| Parameter | $k_{21}$ | $\mu_2$ | $\mu_1$ | $\hat{\mu}_1$ | $\epsilon_1$ | $k_{23}$ | $\epsilon_2$ | $k_{13}$ | $\mu_3$ |
|---|---|---|---|---|---|---|---|---|---|
| $S_T$ index | 4.51e-11 | 1.07e-09 | 7.85e-03 | 1.37e-06 | 1.34e-16 | 2.06e-05 | 8.79e-04 | 2.33e-10 | 2.00e-05 |
| Conf Int | 1.05e-12 | 2.36e-11 | 2.36e-04 | 5.90e-08 | 5.95e-17 | 2.00e-06 | 2.00e-06 | 4.16e-11 | 3.01e-06 |

**Fig. S4:** Total Sobol sensitivities for $N = 2^{15}$, for different initial monomer concentrations ($M(0)$)

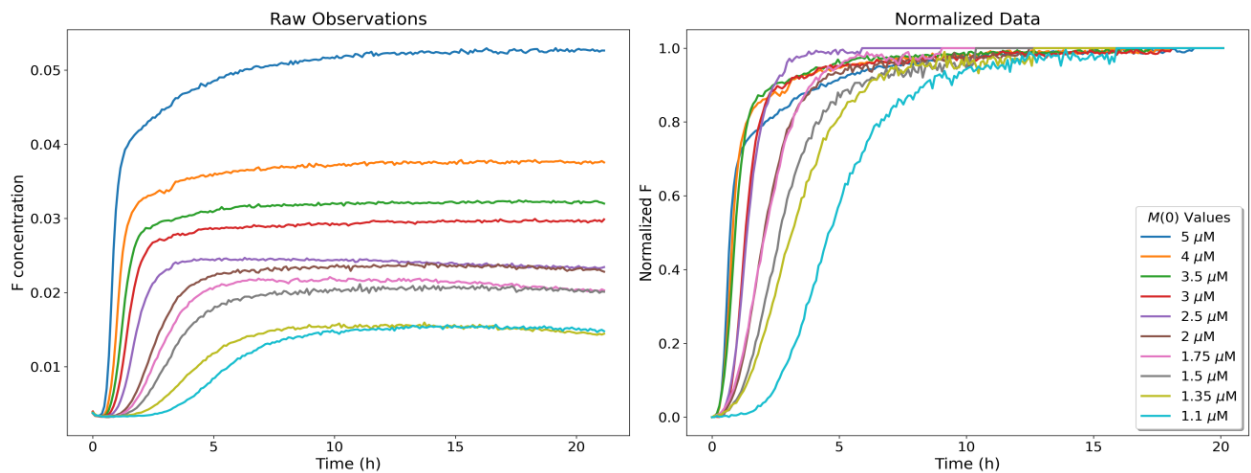

**Fig. S5:** Raw experimental data from (Meisl et al. 2016) for the model output $F$, alongside the normalized $F$, representing the longitudinal concentrations of $F(t)$ with 10 different initial conditions $M(0)$.

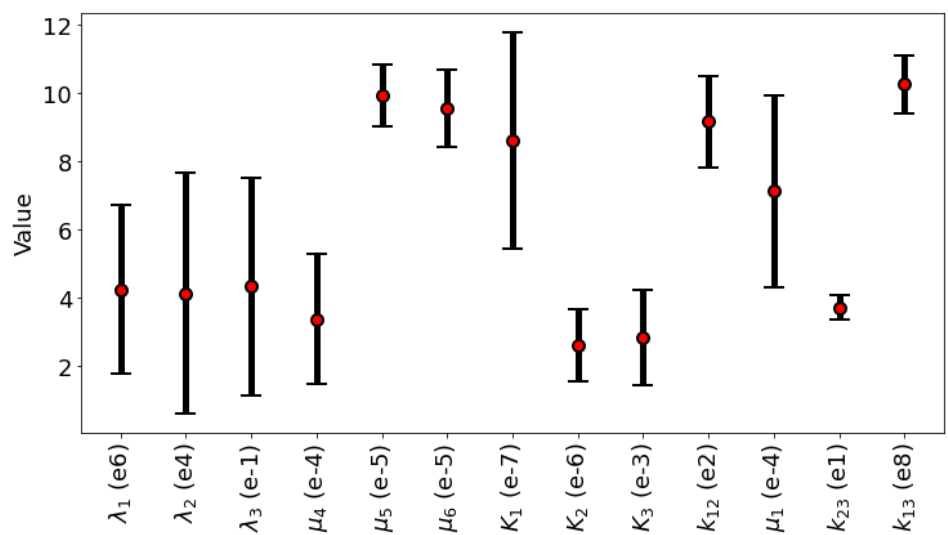

**Fig. S6:** Statistical distribution of the optimized parameter values across different initial monomer concentrations ($M(0)$). Red points indicate the mean of the optimized parameters ($\tilde{\theta}$) across the 10 conditions, with error bars representing standard deviations. Numerical values in parentheses specify the means and standard deviations for each parameter.

**Table S2**: Parameter values $\tilde{\theta}$ for ten different initial concentrations of $M(0)$.

| | $M(0) = 5\mu M$ | $M(0) = 4\mu M$ | $M(0) = 3.5\mu M$ | $M(0) = 3\mu M$ | $M(0) = 2.5\mu M$ | $M(0) = 2\mu M$ | $M(0) = 1.75\mu M$ | $M(0) = 1.5\mu M$ | $M(0) = 1.35\mu M$ | $M(0) = 1.1\mu M$ |
|---|---|---|---|---|---|---|---|---|---|---|
| $\lambda_1$ | $5.5 \times 10^6$ | $4.2 \times 10^6$ | $5.6 \times 10^6$ | $5.0 \times 10^6$ | $1.0 \times 10^6$ | $0.5 \times 10^6$ | $7.5 \times 10^6$ | $6.5 \times 10^6$ | $6.2 \times 10^6$ | $0.5 \times 10^6$ |
| $\lambda_2$ | $5.6 \times 10^3$ | $8.1 \times 10^3$ | $5.8 \times 10^3$ | $9.1 \times 10^3$ | $10.0 \times 10^3$ | $6.6 \times 10^4$ | $8.6 \times 10^4$ | $6.3 \times 10^4$ | $9.9 \times 10^4$ | $6.1 \times 10^4$ |
| $\lambda_3$ | 0.29 | 0.34 | 0.35 | 0.23 | 0.97 | 0.90 | 0.27 | 0.03 | 0.11 | 0.80 |
| $\mu_4$ | $4.1 \times 10^{-4}$ | $3.2 \times 10^{-4}$ | $2.8 \times 10^{-4}$ | $1.8 \times 10^{-4}$ | $1.1 \times 10^{-4}$ | $6.2 \times 10^{-4}$ | $6.3 \times 10^{-4}$ | $5.4 \times 10^{-4}$ | $2.0 \times 10^{-4}$ | $1.0 \times 10^{-4}$ |
| $\mu_5$ | $9.6 \times 10^{-5}$ | $1.1 \times 10^{-4}$ | $9.8 \times 10^{-5}$ | $1.0 \times 10^{-4}$ | $9.1 \times 10^{-5}$ | $1.2 \times 10^{-4}$ | $8.6 \times 10^{-5}$ | $8.9 \times 10^{-5}$ | $1.0 \times 10^{-4}$ | $1.1 \times 10^{-4}$ |
| $\mu_6$ | $6.9 \times 10^{-5}$ | $9.8 \times 10^{-5}$ | $9.2 \times 10^{-5}$ | $1.0 \times 10^{-4}$ | $1.0 \times 10^{-4}$ | $1.1 \times 10^{-4}$ | $8.2 \times 10^{-5}$ | $9.8 \times 10^{-5}$ | $1.1 \times 10^{-4}$ | $9.9 \times 10^{-5}$ |
| $K_1$ | $1.3 \times 10^{-6}$ | $1.2 \times 10^{-6}$ | $9.3 \times 10^{-7}$ | $9.0 \times 10^{-7}$ | $8.6 \times 10^{-7}$ | $9.5 \times 10^{-7}$ | $5.1 \times 10^{-7}$ | $1.2 \times 10^{-6}$ | $4.3 \times 10^{-7}$ | $3.4 \times 10^{-7}$ |
| $K_2$ | $4.0 \times 10^{-6}$ | $3.5 \times 10^{-6}$ | $3.2 \times 10^{-6}$ | $3.7 \times 10^{-6}$ | $3.7 \times 10^{-6}$ | $2.0 \times 10^{-6}$ | $2.0 \times 10^{-6}$ | $1.1 \times 10^{-6}$ | $1.0 \times 10^{-6}$ | $1.9 \times 10^{-6}$ |
| $K_3$ | $3.3 \times 10^{-3}$ | $3.8 \times 10^{-3}$ | $2.2 \times 10^{-3}$ | $4.6 \times 10^{-3}$ | $1.0 \times 10^{-3}$ | $1.1 \times 10^{-3}$ | $1.0 \times 10^{-3}$ | $4.1 \times 10^{-3}$ | $2.7 \times 10^{-3}$ | $4.6 \times 10^{-3}$ |
| $k_{12}$ | $9.9 \times 10^2$ | $1.1 \times 10^3$ | $8.7 \times 10^2$ | $8.2 \times 10^2$ | $1.0 \times 10^3$ | $1.1 \times 10^3$ | $8.0 \times 10^2$ | $8.0 \times 10^2$ | $8.0 \times 10^2$ | $8.0 \times 10^2$ |
| $k_{23}$ | 33.2 | 38.5 | 35.6 | 32.8 | 38.5 | 33.1 | 39.4 | 36.7 | 42.0 | 43.0 |
| $k_{13}$ | $1.0 \times 10^9$ | $1.0 \times 10^9$ | $1.0 \times 10^9$ | $9.1 \times 10^8$ | $1.2 \times 10^9$ | $9.1 \times 10^8$ | $1.0 \times 10^9$ | $1.1 \times 10^9$ | $9.8 \times 10^8$ | $1.1 \times 10^9$ |

**Table S3**: Fixed values of non-sensitive parameters. These parameters were determined based on prior studies or assumed to be negligible within the model's sensitivity range.

| Parameter | $k_{21}$ | $\mu_2$ | $\mu_3$ | $\hat{\mu}_1$ | $\epsilon_1$ | $\epsilon_2$ |
|---|---|---|---|---|---|---|
| Fixed Value | $8 \times 10^{-10}$ | $9 \times 10^{-9}$ | $9 \times 10^{-9}$ | $1 \times 10^{-14}$ | $1 \times 10^{-13}$ | $1 \times 10^{-11}$ |

**Table S4**: Baseline and final amyloid plaque sizes for different Lecanemab treatment plan

| Regimen | 2.5 mg/kg Biweekly | 5 mg/kg Monthly | 5 mg/kg Biweekly | 10 mg/kg Monthly | 10 mg/kg Biweekly |
|---|---|---|---|---|---|
| Baseline | 1.41 | 1.42 | 1.40 | 1.42 | 1.37 |
| Final size | 1.316 | 1.289 | 1.203 | 1.195 | 1.064 |

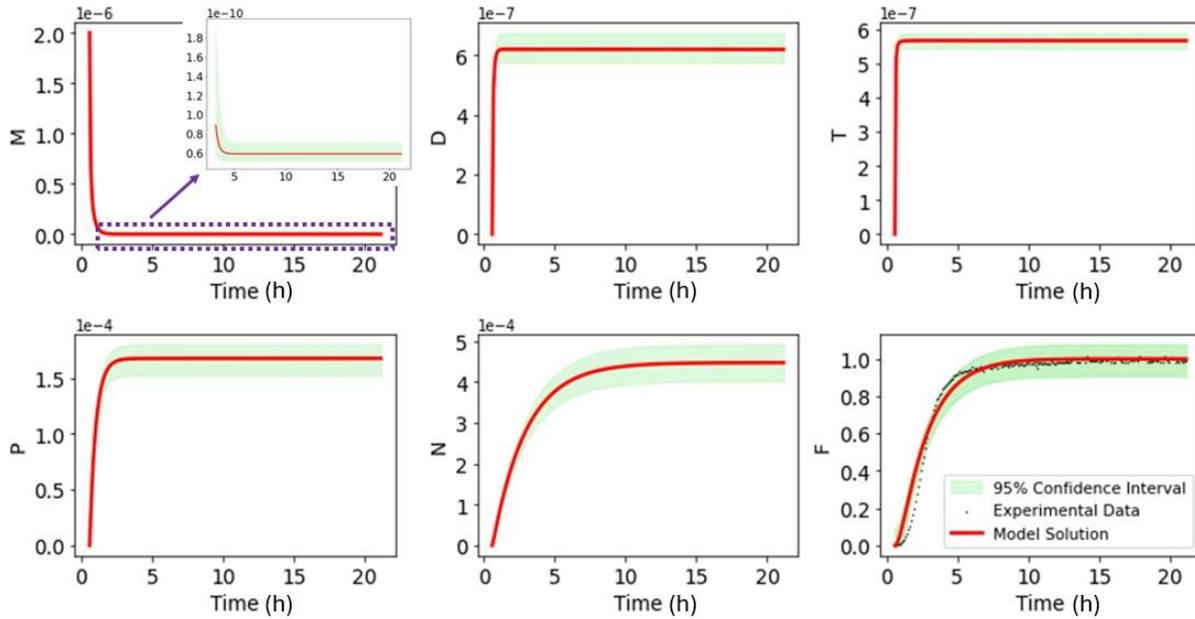

**Figure S7**: Concentration of all state variables with the initial concentration M (0) = $2\mu M$. The shaded areas represent the 95% confidence intervals

**Table S5**: Relationship between $U_{\max}^L$ and $\mu_6$ for each treatment group along with reported ARIA-E

| Treatment Group | $U_{\max}^L$ | ARIA-E (%) |
|---|---|---|
| Lecanemab 2.5 mg/kg biweekly | $\frac{7}{93}\mu_6$ | 1.9 |
| Lecanemab 5 mg/kg monthly | $\frac{9}{91}\mu_6$ | 2.0 |
| Lecanemab 5 mg/kg biweekly | $\frac{14}{86}\mu_6$ | 3.3 |
| Lecanemab 10 mg/kg monthly | $\frac{16}{84}\mu_6$ | 9.9 |
| Lecanemab 10 mg/kg biweekly | $\frac{22}{78}\mu_6$ | 9.9 |

**Table S6**: Baseline and Final Values for Aducanumab Treatment in the Emerge and Engage Groups.

| Group | Low Dose Engage | High Dose Engage | Low Dose Emerge | High Dose Emerge |
|---|---|---|---|---|
| Baseline | 1.385 | 1.407 | 1.394 | 1.383 |
| Final size | 1.215 | 1.175 | 1.224 | 1.105 |
| ARIA-E (%) | 26 | 36 | 26 | 35 |

**Table S7**: Estimated $U_{max}^A$ and ARIA-E for Aducanumab.

| Treatment Group | $U_{max}^A$ ($\mu_6$) | ARIA-E (%) |
|---|---|---|
| Low Dose | $\frac{12}{88}\mu_6$ | 26 |
| High Dose | $0.227\mu_6$ | 35.5 |

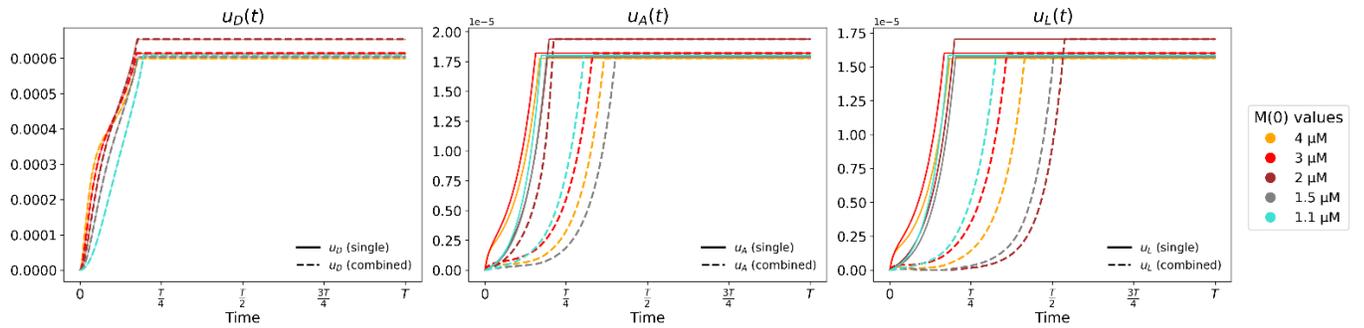

**Figure S8**: Optimal control function values for different initial monomers $M(0)$, comparing single- drug treatments (Left: Donanemab, $u_D$; Center: Aducanumab, $u_A$); Right: Lecanemab, $u_L$ with those obtained from combined therapy scenarios.